\documentclass[12pt]{article}

\usepackage[nosort]{scicite}

\usepackage{times}

\usepackage{longtable}
\usepackage{hyperref}

\newenvironment{sciabstract}{%
\begin{quote} \bf}
{\end{quote}}

\usepackage{graphicx,amsmath,amssymb,geometry,pdflscape}
\title{A characteristic optical variability timescale in astrophysical accretion disks}

\usepackage{authblk}

\author
{Colin J.~Burke$^{1,2}$, Yue Shen$^{1,3\ast}$, Omer Blaes$^{4}$, Charles F.~Gammie$^{1,3,5,6}$, Keith Horne$^{7}$, Yan-Fei Jiang$^8$, Xin Liu$^{1,3}$, Ian M.~McHardy$^9$, Christopher W.~Morgan$^{10}$, Simone Scaringi$^{11}$, Qian Yang$^{1,3}$\\
\normalsize{$^1$Department of Astronomy, University of Illinois at Urbana-Champaign, Urbana, IL 61801, USA}\\
\normalsize{$^2$Center for AstroPhysical Surveys, University of Illinois at Urbana-Champaign, Urbana, IL 61801, USA}\\
\normalsize{$^3$National Center for Supercomputing Applications, University of Illinois at Urbana-Champaign, Urbana, IL 61801, USA}\\
\normalsize{$^4$Department of Physics, University of California, Santa Barbara, CA 93106, USA}\\
\normalsize{$^5$Department of Physics, University of Illinois at Urbana-Champaign, Urbana, IL 61801, USA}\\
\normalsize{$^6$Illinois Center for Advanced Study of the Universe, University of Illinois at Urbana-Champaign, Urbana, IL 61801, USA}\\
\normalsize{$^7$School of Physics and Astronomy, University of St Andrews, Fife, KY16 9SS, UK}\\
\normalsize{$^8$Center for Computational Astrophysics, Flatiron Institute, New York, NY 10010, USA}\\
\normalsize{$^9$Department of Physics \& Astronomy, University of Southampton, Southampton SO17 1BJ, UK}\\
\normalsize{$^{10}$Department of Physics, United States Naval Academy, Annapolis, MD 21402, USA}\\
\normalsize{$^{11}$Department of Physics, University of Durham, Durham DH1 3LE, UK}\\
\normalsize{$^\ast$To whom correspondence should be addressed; E-mail:  shenyue@illinois.edu.}
}

\newcommand{\etal}{et al.}

\usepackage{color,subfigure}

\def\apj{Astrophys. J.}
\def\apjl{Astrophys. J.}
\def\aj{Astron. J.}
\def\mnras{Mon. Not. R. Astron. Soc.}
\def\apjs{Astrophys. J. Suppl. Ser.}
\def\araa{Annu. Rev. Astron. Astrophys.}
\def\aap{Astron. Astrophys.}

\date{}

\begin{document}


\baselineskip24pt


\maketitle


\begin{sciabstract}
Accretion disks around supermassive black holes in active galactic nuclei produce continuum radiation at ultraviolet and optical wavelengths. Physical processes in the accretion flow lead to stochastic variability of this emission on a wide range of timescales. We measure the optical continuum variability observed in 67 active galactic nuclei and the characteristic timescale at which the variability power spectrum flattens. We find a correlation between this timescale and the black hole mass, extending over the entire mass range of supermassive black holes. This timescale is consistent with the expected thermal timescale at the ultraviolet-emitting radius in standard accretion disk theory. Accreting white dwarfs lie close to this correlation, suggesting a common process for all accretion disks.  
\end{sciabstract}

% In setting up this template for *Science* papers, we've used both
% the \section* command and the \paragraph* command for topical
% divisions.  Which you use will of course depend on the type of paper
% you're writing.  Review Articles tend to have displayed headings, for
% which \section* is more appropriate; Research Articles, when they have
% formal topical divisions at all, tend to signal them with bold text
% that runs into the paragraph, for which \paragraph* is the right
% choice.  Either way, use the asterisk (*) modifier, as shown, to

%\section*{Main Text:}

Accretion disks are present around growing supermassive black holes (SMBHs) found in active galactic nuclei (AGNs). Standard theory of radiatively-efficient accretion disks\cite{SSD} can reproduce the broad-band emission from AGNs\cite{Shields,Sun_Malkan}, but the exact structure and physical processes occurring in accretion disks remain unknown. Because AGN accretion disks are too small to resolve in direct observations, constraints on their structure have been derived from gravitational microlensing\cite{Morgan10,Morgan18} and time delay measurements of accretion disk echos to flux variations from the innermost region around the SMBH\cite{Sergeev05, Cackett07,Fausnaugh,Edelson17}. 

Optical emission from AGN accretion disks exhibit stochastic variability, for unknown reasons\cite{Ulrich97,Padovani17}. Optical light curves (i.e., time series of fluxes tracing the variable accretion disk emission) for large samples of AGNs can be used to measure the variability characteristics of the accretion disk emission. The power spectrum density (PSD) of AGN optical variability can be approximated by a Damped Random Walk (DRW) model\cite{kelly09,koz10,MacLeod10,Suberlak21,Zu13,Simm16}, with a $f^{-2}$ power law ($f$ is the frequency) at the high-frequency end and white noise at the low-frequency end\cite{kelly09,koz10}. There are deviations from the $f^{-2}$ scaling at the highest frequencies\cite{Zu13,Mushotzky} in some individual AGNs. The transition frequency, corresponding to a characteristic damping timescale $\tau_{\rm damping}$, is typically several hundred days for quasars\cite{MacLeod10,Suberlak21,Simm16}, the most luminous subset of AGNs with a bolometric luminosity $L_{\rm bol}\gtrsim 10^{45}\,{\rm erg\,s^{-1}}$. There is no widely accepted physical interpretation for this damping timescale. There is tentative evidence that this damping timescale may correlate with the mass of the SMBH and/or luminosity of the AGN\cite{kelly09,MacLeod10,Collier_Peterson01} but such claims have been controversial\cite{Simm16} and the results inconsistent\cite{kelly09,MacLeod10,Suberlak21}. The range of SMBH mass in those studies has been limited to two orders of magnitude, and the measurements of the damping timescales are susceptible to biases due to the limited observing period\cite{koz17,methods}. 

To address these limitations, we compiled optical light curves from the literature for AGNs with estimated SMBH masses. We excluded any light curves that did not have sufficient signal-to-noise ratio or duration to robustly constrain the damping timescale\cite{methods}. Starting from an initial set of $\sim 400$ AGNs, our selection criteria lead to a final sample containing $67$ AGNs that span the entire SMBH mass range of $\sim 10^4-10^{10}$ solar masses ($M_\odot$). Throughout this paper, all timescales were converted to the rest-frame of the AGN and all quoted uncertainties and scatter are $1\sigma$ unless otherwise specified.

\begin{figure}[!t]
\centering
\includegraphics[width=1.0\textwidth]{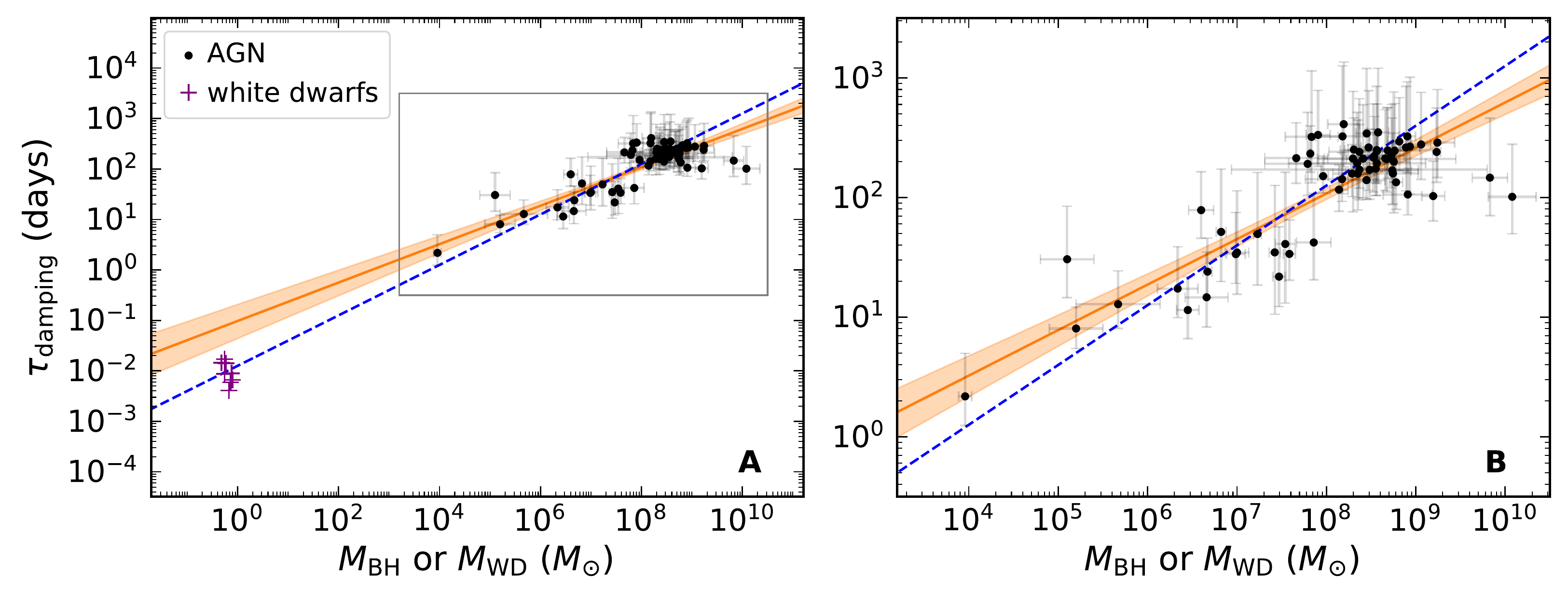}
\caption{\textbf{The optical variability damping timescale as a function of accretor mass.} The region in the grey box in Panel A is shown in Panel B. The rest-frame damping timescale $\tau_{\rm{damping}}$ was measured from AGN light curves and correlates with SMBH mass $M_{\rm BH}$ for AGNs (black circles). The orange line and shaded band are the best-fitting model and $1\sigma$ uncertainty for the AGN sample. Purple crosses show equivalent measurements for white dwarfs\cite{Scaringi15}, where $M_{\rm WD}$ denotes the mass of the white dwarf; these do not fall in the orange band but are consistent with a model that has a fixed mass slope of 0.5 (blue dashed line). The typical uncertainties on $M_{\rm WD}$ and the white dwarf damping timescale are 0.2 dex and 0.01 days respectively \cite{Scaringi15}. All error bars are $1\sigma$. } \label{fig:tau_mass}
\end{figure}

\begin{figure}[!h]
\centering
\includegraphics[width=1.0\textwidth]{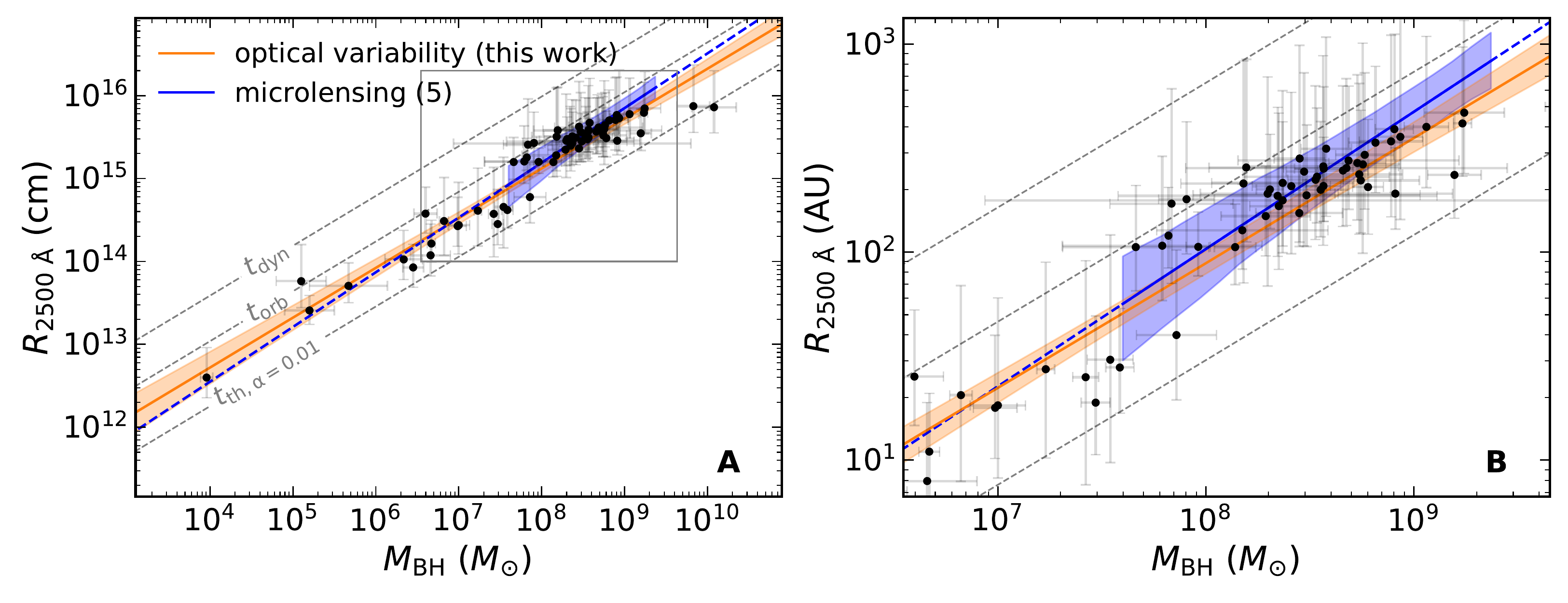}
\caption{\textbf{The accretion disk emitting radius at rest-frame 2500\,\AA\ as a function of SMBH mass.} The region in the grey box in Panel A is shown in Panel B. The emitting radius is computed as $R_{2500\,\textrm{\AA}} \propto M_{\rm{BH}}^{1/3}\alpha^{2/3}\tau_{\rm{damping}}^{2/3}$, assuming $\tau_{\rm damping}$ is the thermal time and $\alpha=0.05$. The data points (black circles) are overplotted with the best-fitting linear model (orange line) and $1\sigma$ uncertainty (orange shaded area). The relation derived from microlensing observations\cite{Morgan18} is shown in the overlapping mass range (blue solid line and $1\sigma$ shaded region). The blue dashed line indicates an extrapolation of the microlensing results to other black hole masses. The three gray dashed lines are the corresponding radius if $\tau_{\rm damping}$ is identified as $t_{\rm dyn}$, $t_{\rm orb}$, or $t_{\rm th}$ with $\alpha=0.01$, respectively. All error bars are $1\sigma$.} \label{fig:radius_mass}
\end{figure}

Figure~\ref{fig:tau_mass} shows the relation between our derived damping timescales and SMBH masses. There is a correlation (Pearson correlation coefficient $r=0.82$) over the SMBH mass range of $\sim 10^4-10^{10}\,M_\odot$. We verified that this correlation persists if we make different choices for the details in measuring the damping timescale or methods of SMBH mass estimation\cite{methods}. The best-fitting model relation is:
\begin{equation}\label{eqn:tau_mass}
    \tau_{\rm{damping}} = 107^{+11}_{-12}\ {\rm{days}}\  \left(\frac{M_{\rm{BH}}}{10^8\ M_{\odot}}\right)^{0.38^{+0.05
    }_{-0.04}}\ ,
\end{equation}
where $M_{\rm BH}$ is the mass of the SMBH. The data have an additional $1\sigma$ intrinsic scatter of $0.09^{+0.05}_{-0.04}$ dex around the best-fitting model. This relation is sufficiently tight that inversion of Eqn.~(1) can predict SMBH mass given $\tau_{\rm damping}$ with a $1\sigma$ precision of $\sim 0.3\,$dex. Alternatively, fitting a linear model for $M_{\rm BH}$ given $\tau_{\rm damping}$ yields 
\begin{equation}\label{eqn:mass_tau}
M_{\rm{BH}} = 10^{7.97^{+0.14}_{-0.14}}\ M_{\odot}\ \left(\tau_{\rm{damping}} / 100\  {\rm{days}} \right)^{2.54^{+0.34}_{-0.35}}, 
\end{equation}
with an intrinsic scatter of $0.33^{+0.11}_{-0.11}$ dex in $M_{\rm{BH}}$. This intrinsic scatter in the predicted SMBH mass is similar to the systematic uncertainties in SMBH mass measurements\cite{Peterson,Shen13}. Previous studies of AGN optical variability have found that the damping timescale depends weakly on wavelength $\lambda$\cite{MacLeod10,Suberlak21} as $\tau_{\rm damping}\propto \lambda^{0.17}$. Figure~\ref{fig:radius_mass} utilizes the measured damping timescale in different bands (and at different redshifts) scaled using this relation to rest-frame wavelength $2500$\,\AA. Because lower-mass systems are generally at lower redshifts than higher-mass systems in our AGN sample, due to observational biases, a positive wavelength dependence of $\tau_{\rm damping}$ slightly flattens its observed mass dependence. However, the measured weak wavelength dependence of the damping timescale\cite{MacLeod10,Suberlak21} means the mass dependence (i.e., the slope of the $\tau_{\rm damping}-M_{\rm BH}$ relation) at a fixed rest-frame wavelength is still below 0.5. 

The light curve duration of our dataset is generally not long enough to constrain the damping timescale in the most massive ($>10^9\,M_\odot$) or distant ($z>1$) AGNs. By restricting our sample to AGNs with $\tau_{\rm damping}$ shorter than one tenth of the light curve baseline\cite{methods}, we potentially introduce a bias by underestimating the average damping timescale for the most massive or distant AGNs. Nevertheless, this caveat does not affect the existence of a damping timescale -- mass correlation\cite{methods}.  

Most radiatively-efficient AGNs accrete within a narrow range of accretion rates (normalized by SMBH mass), with an average Eddington ratio (the ratio between the total luminosity $L_{\rm bol}$ and a mass-dependent characteristic luminosity $L_{\rm Edd}\equiv 1.3\times 10^{38}(M_{\rm BH}/M_\odot)\,{\rm erg\,s^{-1}}$) $L_{\rm bol}/L_{\rm Edd}\sim 0.2$ and a dispersion of $\sim 0.3$\,dex \cite{Kollmeier06,Shen11}, similar to the median and dispersion of Eddington ratios in our sample\cite{methods}. We find no correlation between the model residuals and $L_{\rm bol}/L_{\rm Edd}$, which we interpret as due to the limited dynamic range and large systematic uncertainties in the measured Eddington ratios. Any dispersion in the true $L_{\rm bol}/L_{\rm Edd}$ (that is, free of measurement uncertainties) can potentially contribute to the intrinsic scatter around the average relation.

To extend the $\tau_{\rm damping}$-mass scaling relation to accretors with much smaller masses, we consider accreting white dwarfs that are non-eruptive (that is, the accretion rate is approximately stable). The $\tau_{\rm damping}$ measurements for white dwarfs are based on optical light curves and taken directly from \cite{Scaringi15}. A $\tau_{\rm damping}$-mass scaling with a mass slope of 0.5 is consistent with measurements for accretion disks in these white dwarfs (Figure~\ref{fig:tau_mass} and \cite{methods}). We do not consider optical variability in accretion disks around neutron stars or stellar-mass black holes because the optical accretion disk emission could be complicated by X-ray reprocessing\cite{Done}, or simply overwhelmed by optical light from a companion star.  

This average scaling between the damping timescale and SMBH mass can be qualitatively understood within the standard theory of accretion disks. Both the orbital time $t_{\rm orb}$ (the time to orbit around the black hole) and the thermal time $t_{\rm th}$ (the timescale to restore thermal equilibrium) scale with the SMBH mass and radius as\cite{SSD}:
\begin{eqnarray}\label{eqn:eq2}
t_{\rm orb}=100\left(\frac{M_{\rm BH}}{10^8\,M_\odot}\right)\left(\frac{R}{100R_S}\right)^{3/2}\, {\rm days}\ ,\\
t_{\rm th}=1680\left(\frac{\alpha}{0.01}\right)^{-1}\times\left(\frac{M_{\rm BH}}{10^8\,M_\odot}\right)\left(\frac{R}{100R_S}\right)^{3/2}\, {\rm days}\ ,
\end{eqnarray}
where $R_S=2GM_{\rm BH}/c^2$ is the Schwarzschild radius of the black hole ($G$ is the gravitational constant and $c$ is the speed of light in vacuum), and $\alpha$ is the viscosity parameter. The relation among different timescales is: $t_{\rm th}\approx (2\pi\alpha)^{-1}t_{\rm orb}=\alpha^{-1}t_{\rm dyn}\approx (H/R)^2t_{\rm vis}$, where $t_{\rm dyn}$ is the dynamical time (an alternative to $t_{\rm orb}$ used in the literature), $t_{\rm vis}$ is the viscous time on which matter diffuses through the accretion disk due to viscosity, and $H/R$ is the ratio between the scale height $H$ and the radial extent $R$ of the accretion disk. 

Assuming all AGNs accrete at constant Eddington ratio ($\dot{M}_{\rm BH}/M_{\rm BH}\propto L_{\rm bol}/L_{\rm Edd}=$ constant, where $\dot{M}_{\rm BH}$ is the mass accretion rate) and any dispersion in Eddington ratio leads to intrinsic scatter around the average relation, the standard theory predicts a scaling relation between the effective emitting radius (at a given rest wavelength) and the black hole mass as $R_{\lambda}\propto M_{\rm BH}^{2/3}$. We assume that the radiative efficiency $\eta\propto L_{\rm bol}/\dot{M}_{\rm BH}$ is also constant. In reality, our sample contains AGNs with different accretion rates and possibly a range of black hole spins, which may lead to different values of $\eta$, introducing additional scatter around the average relation. At a given rest-frame wavelength, the orbital and thermal timescales therefore scale with mass as $t_{\rm orb}, t_{\rm th}\propto M_{\rm BH}^{1/2}$. If the damping timescale we measure is associated with the orbital time or the thermal time, then we expect a slope of 0.5 in the $\tau_{\rm damping}-M_{\rm BH}$ relation, which is consistent with the observed slope within $2.5\sigma$. A steeper wavelength dependence of $\tau_{\rm damping}$ than previously reported\cite{MacLeod10}, as discussed above, would improve the agreement.   

The physical origin of the damping timescale could
be associated with the thermal timescale at the radius where variability is driven. To compare our results to AGN disk sizes measured from microlensing, we first scale the damping timescale to rest-frame 2500\,\AA\ using its measured wavelength dependence $\tau_{\rm damping}\propto \lambda^{0.17}$\cite{MacLeod10}. Assuming this ultraviolet (UV)-emitting part of the disk is where variability is driven, we derive the effective UV-emitting radius as $R_{2500\,\textrm{\AA}}\propto M_{\rm BH}^{1/3}\alpha^{2/3}\tau_{\rm damping}^{2/3}$ using Eqn.~(4). 

Figure~\ref{fig:radius_mass} shows the relation between the derived physical radius $R_{2500\,\textrm{\AA}}$ that emits at rest-frame 2500\,\AA\ and the SMBH mass for our sample. We find a correlation that is consistent with the prediction from the standard model, $R_{2500\,\textrm{\AA}}\propto M_{\rm BH}^{2/3}$. We have assumed a fiducial viscosity parameter of $\alpha=0.05$, which leads to $t_{\rm th}\approx 3t_{\rm orb}\approx 20t_{\rm dyn}$. This fidicual value of $\alpha$ is higher than the typical value of $\sim 0.01$ found in standard magnetohydrodynamic (MHD) simulations of accretion disks\cite{MRI}, but is consistent with the range of $0.05-0.1$ in simulations of radiation-pressure-dominated AGN accretion disks (see supplementary text). Figure~2 also shows the relation derived from microlensing measurements of accretion disk sizes for luminous quasars\cite{Morgan18}. 

The size-mass relation we derived from the optical variability data is consistent with the constraints from microlensing in the overlapping mass range (Figure~2), but extends to lower masses. This suggests association of the damping timescale with the thermal time at the effective UV-emitting radius. Because the normalization of our $\tau_{\rm damping}-M_{\rm BH}$ relation is constrained to within $\sim 10\%$ (1$\sigma$) and we consider the microlensing results reliable, $t_{\rm dyn}$ or $t_{\rm orb}$ is less favorable than $t_{\rm th}$ (with $\alpha=0.05$) as the origin for $\tau_{\rm damping}$. If $\tau_{\rm damping}$ is associated with $t_{\rm th}$, $\alpha$ must be in the range $0.03-0.12$ to be within $\sim1\sigma$ of the microlensing constraints. Both our analysis and the microlensing study assume the same standard accretion disk model, but differ in additional assumptions. For example, our variability approach assumes that the damping timescale is the thermal timescale with a fiducial viscosity parameter $\alpha=0.05$, without needing to know the orientation of the disk; the microlensing analysis did not make this assumption on $\alpha$ but assumed a mean orientation of the disk. The correlation in Figure~\ref{fig:radius_mass} is tighter (Pearson correlation coefficient $r=0.96$) than in Figure~\ref{fig:tau_mass} because the computation of $R_{2500\,\textrm{\AA}}$ includes an explicit mass dependence, biasing the correlation strength. Nevertheless, the agreement with the microlensing results at the high-mass end and a different scaling than $R_{2500\,\textrm{\AA}} \propto M_{\rm BH}^{1/3}$ (expected from pure self-correlation) lead us to conclude that the correlation seen in Figure~\ref{fig:radius_mass} is not due to self-correlation. 

%The limited range of $L_{\rm bol}/L_{\rm Edd}$ for broad-line AGN and the $\tau_{\rm damping}-M_{\rm BH}$ relation imply a relation between the damping timescale and the optical luminosity of the disk. This relation is also tight for our AGN sample: $\tau_{\rm{damping}} = 10^{1.92^{+0.05}_{-0.05}}\ {\rm{days}}\ (L_{\rm{5100}\AA}/10^{44}\ \rm{erg\ s}^{-1})^{0.33^{+0.04}_{-0.04}}$ with an intrinsic scatter of $0.11^{+0.05}_{-0.05}$ dex. If we compute the rest-frame 2500\,\AA\ radius as before and correlate with the optical continuum luminosity, we derive a scaling relation $R_{\rm 2500\,\AA}\propto L_{5100\,\AA}^{0.53\pm0.04}$, consistent with the expectation $R_{\rm 2500\,\AA}\propto L_{\nu}^{0.5}$ from the standard model\cite{methods,Davis_Laor}.

Figure~3 compares the optical damping timescale of the accretion disk to the characteristic X-ray variability timescale at different AGN SMBH masses, where the X-ray variability measurements were taken from \cite{xray}. X-ray emission in AGNs mainly arises from a hot optically-thin but geometrically-thick gas (a region commonly referred to as the ``corona''), much closer to the SMBH than the UV-optical part of the accretion disk, and the characteristic X-ray variability timescales are expected to be substantially shorter (e.g., Eqn.~3 and Eqn.~4). The best-fitting timescale-mass relation for X-ray variability has a slope close to unity\cite{Scaringi15,xray,McHardy06,Kording07}, as expected if the X-ray timescale traces the orbital or thermal time near the innermost stable circular orbit (ISCO), which has a radius that scales linearly with black hole mass. In contrast, the characteristic timescales measured from optical variability are several orders of magnitude longer than the X-ray timescale, and have a different mass dependence. The higher scatter in the X-ray relation than in the optical relation could be due to other parameters that affect the ISCO radius (such as black hole spin), or by the range of X-ray-emitting radii within the optically-thin corona. The thermal timescale described by Eqn.~(4) does not apply to the corona, but we expect that any X-ray variability would operate on the orbital timescale in the corona region.  

The correlation between the damping timescale of optical variability and the SMBH mass has implications for AGN accretion disk models. It implies an intrinsic origin for AGN optical variability, as opposed to extrinsic causes such as microlensing. The difference with X-ray variability rules out simple reprocessing of X-ray emission into optical \cite{Done} as the origin of the optical variability on similar timescales as the damping timescale, requiring internal accretion disk processes that either drive the optical variability themselves or modify the X-ray reprocessing.

\begin{figure}[!h]
\centering
\includegraphics[width=0.8\textwidth]{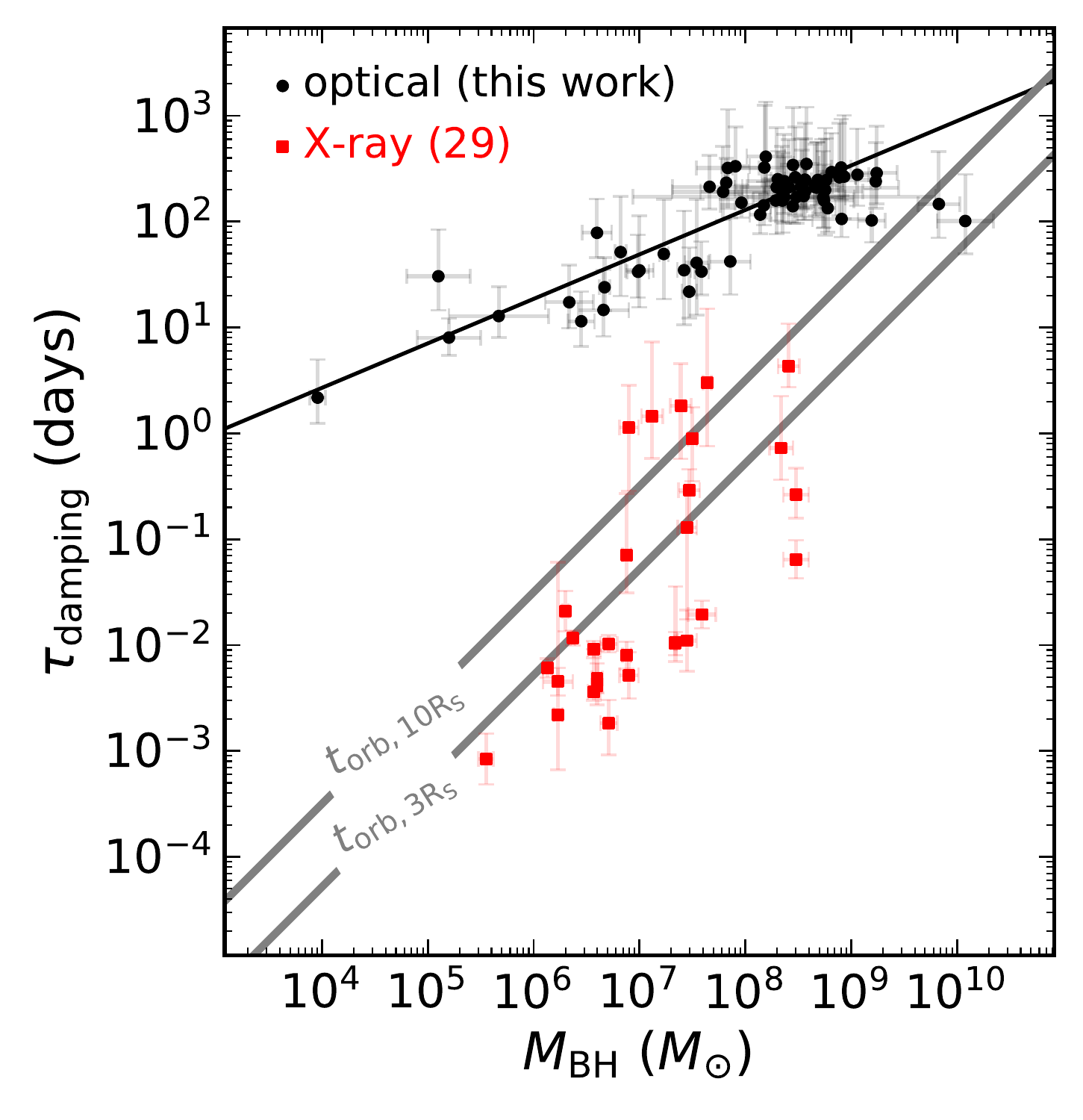}
\caption{\textbf{Comparison of AGN variability timescales at optical and X-ray wavelengths.} Red squares show measurements of the break timescale for X-ray variability\cite{xray} and the black circles are our optical measurements (same as Fig.~1), both as functions of the SMBH mass. The thick gray lines indicate the orbital timescale at ten and three times the Schwarzschild radius, which has a linear dependence on mass. The optical and X-ray data have different slopes. The correlation in the optical is tighter than the X-ray correlation. All error bars are $1\sigma$. } \label{fig:br_mass}
\end{figure}

There is no detailed physical model that can explain the observed variability characteristics of accretion disks (see supplementary text), but the standard accretion disk model\cite{SSD} provides qualitative agreement with the observed timescale-mass relation over ten orders of magnitude in accretor mass (combining AGNs and white dwarfs). The association of the characteristic variability timescale with the thermal timescale of the accretion disk explains the observed low-frequency break in the optical variability PSD. Fig.~\ref{fig:br_mass} and Fig.~\ref{fig:tau_orb} show that the X-ray variability timescale is consistent with the orbital time at the ISCO, although the large scatter and the small number of stellar mass black holes (in Fig.~S7) cannot rule out other timescales (such as the viscous time $t_{\rm vis}$) as the origin for the break in the X-ray variability PSD.

It remains unclear which processes drive these accretion disk flux variations, and whether additional accretion parameters (such as accretion rate, black hole spin, etc.) are involved. The measured AGN accretion disk sizes from microlensing are larger than predicted by the standard model\cite{Dexter,Char}. The measured wavelength dependence of the damping timescale\cite{MacLeod10} is shallower than standard model predictions ($\tau\propto \lambda^2$), implying a more complicated mapping from the damping timescales to the physical radii as a function of wavelength. We speculate that the variability is driven in the inner part of the accretion disk, emitting at rest-frame UV, which induces optical variability by rapid outward propagation, during which the damping timescale approximately preserved (supplementary text). In other words, the damping timescale traces the thermal timescale at the UV-emitting part of the accretion disk, even when measured at longer (e.g., optical) wavelengths.

Regardless of the physical mechanism, the observed $\tau_{\rm damping}-M_{\rm BH}$ relation can be used to estimate the SMBH mass of an AGN using optical variability. The correlation parameters are sufficiently well-constrained to provide mass estimates that are as accurate as reverberation mapping and single-epoch methods. The method can be applied to AGNs at the low-mass end of SMBHs, where the broad-line emission is often too weak to measure a robust SMBH mass using spectral methods.

\clearpage

\clearpage

%, and can serve as a discovery machine to search for elusive accreting black holes in the intermediate ($10\,M_\odot-10^4\,M_\odot$) mass range, \blue{given that the optical emission is dominated by the accretion disk}. 

% Change the title of the references section
\renewcommand\refname{References and Notes}

\bibliographystyle{unsrt}

%% Here is the endmatter stuff: Supplementary Info, etc.
%% Use \item's to separate, default label is "Acknowledgements"

\section*{Acknowledgments}

\textbf{Funding}: C.J.B. acknowledges support from the Illinois Graduate Survey Science Fellowship. Y.S. was supported by NSF grant AST-2009947. C.F.G. was supported by NSF grants AST-1716327 and OISE-1743747. K.H. was supported by UK STFC grant ST/R000824/1. I.M.M. was supported by UK STFC grant ST/R000638/1. C.W.M. was supported by NSF grant AST-2007680. \textbf{Author contributions:} C.J.B. led the data compilation and analysis; Y.S. designed the project and led the manuscript writing; O.B., C.F.G. and Y.-F.J. led the theoretical interpretation; X.L. and Q.Y. contributed to data compilation; I.M.M. and S.S. led the X-ray variability and white dwarf discussion; C.W.M. led the microlensing discussion; K.H. led the disk reverberation mapping discussion; all authors contributed to the science interpretation and manuscript writing. \textbf{Competing interests:} The authors declare that they have no competing interests. \textbf{Data and materials availability:} Optical light curves of the AGN sample were taken from the publicly-available sources listed in Table S1 and Data S1, and we make the compiled optical light curves available online \cite{Zenodo}. White dwarf optical variability timescale measurements were taken from \cite{Scaringi15}. X-ray variability timescale measurements were from \cite{Scaringi15,xray}. Our derived optical $\tau_{\rm damping}$ measurements are provided in Table S1 (for the final sample) and Data S1 (for the initial sample). Software used in this work can be accessed from public code repositories of the cited references, and a python notebook to fully reproduce our analysis is provided at \url{https://github.com/burke86/taufit/tree/master/paper}. The full figure set for our initial AGN sample (an example shown in Fig.~S5) is available at \cite{Zenodo}.

%\section*{Supplementary materials}
%Materials and Methods\\
%Supplementary Text\\
%Figs. S1 to S7\\
%Table S1\\
%Data S1\\
%References \textit{(34--91)}

% Change the table names to e.g. Table S1
\renewcommand{\thetable}{S\arabic{table}}

\clearpage
\setcounter{page}{1}

\section*{Materials and Methods}

\subsection*{Sample and Data}

We collected optical light curves of broad-line AGNs with SMBH mass measurements in the literature. These SMBH masses are either based on reverberation mapping [RM, \cite{Peterson}] or single-epoch (SE) virial SMBH mass methods\cite{Shen13}, with a typical systematic uncertainty of $\sim 0.4$\,dex. Many of the reported measurement uncertainties in SMBH mass for our sample are equal to or smaller than the systematic uncertainty, but we opt to use only the measurement uncertainty in SMBH mass as this provides an upper limit of the intrinsic scatter of the correlations. 

To measure the optical variability of AGNs over a large dynamic range in SMBH mass, we selected our sample to include both distant quasars with decade-long photometric light curves, as well as nearby AGNs targeted for reverberation mapping. The parent AGN sample includes reverberation mapping AGNs (mostly at $z<0.3$) from the AGN Black Hole Mass Database\cite{Bentz}, and $z>0.3$ quasars with light curves from ground-based monitoring\cite{kelly09,Geha03,Yang,Giveon99,Walsh09}. These objects mostly populate the $>10^7\,M_\odot$ mass range. To improve the statistics in the low-mass regime ($M_{\rm BH}< 10^7\,M_\odot$), we supplement our sample with broad-line dwarf AGNs with SMBH masses measured using single-epoch methods\cite{RGG,Chilingarian18,Guo,Cann20}. Finally, we include the SMBH in the dwarf AGN NGC 4395, which has the lowest mass in our sample of $\sim 10^4\,M_\odot$\cite{Woo} and high-cadence optical light curves from the \emph{Transiting Exoplanet Survey Satellite} [TESS, \cite{TESS,Burke20}]. All these objects are broad-line AGNs (non-blazars) for which the optical continuum mainly probes the accretion disk emission. 

The optical light curves for our AGN sample are from several surveys: the compiled 20-year-long photometric light curves for quasars in the Sloan Digital Sky Survey (SDSS) Stripe 82 region \cite{Shen11,Yang} (restricting to those with densely sampled seasonal light curves), light curves for individual AGN from various reverberation mapping programs\cite{Walsh09,Du18,Bentz14,Bentz16a,Bentz16b,Pei14,Barth11,Peterson14,Lu16,Hu20,Fausnaugh17,Bentz09,Kaspi00,Cackett20,Denney10,De_Rosa18,Du14,Pancoast19,peterson04,williams2018}, and light curves from wide-area time-domain imaging surveys\cite{DES,ZTF,CRTS,ASASSN}. Some of the optical light curves were combined by \cite{Yang} from multiple facilities to extend the baseline, incorporating calibration coefficients to account for small differences in bandpasses. These optical light curves sample different rest-frame wavelengths for objects at different redshifts. These wavelength differences are taken into account when computing the accretion disk size below. Because the wavelength correction is mild\cite{MacLeod10,Suberlak21}, this particular detail about wavelength in these light curves will not affect our results.

Our initial sample includes $\sim 400$ AGNs with long-duration (up to $\sim20$ year-long) light curves and more than 5 orders of magnitude in SMBH mass, from which we select a final sample with robust measurements of the damping timescale. The final sample is listed in Table S1, and the initial sample in Data S1, including basic AGN properties collected from the literature and our timing analysis results (see below). When not available from the primary sources, we collected the measurements of the continuum luminosity at rest-frame 5100\,\AA\ $L_{5100\ \textrm{\AA}}$ from other publications\cite{Savic18,Le20,Du19,Cho}. Bolometric luminosities are computed from $L_{5100\ \textrm{\AA}}$ assuming a constant bolometric correction\cite{Shen11} of 9.26. For our final sample, the distribution of Eddington ratios has a median of 0.15 and a dispersion of $\sim 0.3$\,dex.

\subsection*{Timing Analysis}
\label{sec:parameterinference}

The Damped Random Walk (also known as the Ornstein--Uhlenbeck process) models the stochastic optical AGN variability which traces the accretion disk emission\cite{kelly09,koz10,MacLeod10,Suberlak21}, although there have been reported deviations in the high-frequency end\cite{Zu13,Mushotzky} such that the slope of the PSD becomes steeper than $-2$. This does not affect our results because we measure the transition to a white noise spectrum at the low-frequency end. Some optical variability measurements also found a low-frequency slope close to $-1$ instead of a white noise slope\cite{Simm16}. However, this may be due to insufficient duration of the light curves, biasing the slope measurements in the low-frequency regime. As we demonstrate below, a forced fit of a DRW model can still recover the damping (transition) timescale in such cases.

The DRW model is the simplest case of a family of continuous autoregressive with moving average (CARMA) models\cite{kelly14} for Gaussian processes. Unlike most of the higher-order CARMA models, the DRW model provides an interpretation of the curvature in the PSD as being due to the damping of power beyond a characteristic timescale. For this reason, we adopt the DRW model as our fiducial model to measure the characteristic break timescale/frequency in the PSD, but we acknowledge that the actual variability process can be more complicated than a DRW. In our DRW analysis, the conversion between timescale $\tau$ and frequency $f$ is $\tau=1/(2\pi f)$. For convenience of the discussion, we use damping timescale or characteristic timescale to refer to the inflection in the PSD, which is often referred to as the break or bend timescale/frequency in X-ray timing studies. 

We use Gaussian process (GP) regression to fit a DRW model to the light curves as implemented in the \textsc{celerite} package\cite{Foreman-Mackey17}, which produces consistent results as those using alternative packages such as the \textsc{carma\_pack}\cite{kelly14}. A GP model is defined completely by its covariance function (or kernel function). The covariance function for a DRW is
\begin{equation}
k(t_{nm}) = 2 \sigma_{\rm{DRW}}^2 \exp{(-t_{nm} / \tau_{\rm{DRW}})}\ \tag{S1},
\end{equation}
where $t_{nm}=|t_{n}-t_{m}|$ is the time lag between measurements $m$ and $n$, $\sigma_{\rm{DRW}}$ is the amplitude term, and $\tau_{\rm{DRW}}$ is the damping timescale. This has an exact correspondence to the structure function (SF) definition,

%\begin{equation}
\begin{align}
    {\rm{SF}}^2 & = \rm{SF}_{\infty}^2(1 - {\rm{ACF}}(t_{nm})) \tag{S2}\\
    & = 2 \sigma_{\rm{DRW}}^2 (1 - \exp{(-t_{nm} / \tau_{\rm{DRW}}}))\ , \tag{S3}
    \label{eq:drw}
\end{align}
%\end{equation}
where the asymptotic variability amplitude $\rm{SF}_{\infty} = \sqrt{2} \sigma_{\rm{DRW}}$ and the autocorrelation function ${\rm{ACF}}(t_{nm}) = \exp{(-t_{nm} / \tau_{\rm{DRW}}})$.

Light curves often include white noise in excess of the quoted measurement errors. We performed simulations to test whether excess white noise can bias $\tau_{\rm{DRW}}$ to smaller values. To address this, we add an excess white noise term and derive the final kernel function as:
\begin{equation}
    k(t_{nm}) = 2 \sigma_{\rm{DRW}}^2 \exp{(-t_{nm} / \tau_{\rm{DRW}})} + \sigma_{n}^2\delta_{nm}\ , \tag{S4}
    \label{eq:drw}
\end{equation}
where $\sigma_{\rm{n}}$ is the excess white noise amplitude, and $\delta_{nm}$ is the Kronecker $\delta$ function. Therefore, we have three free parameters to fit in the model: $\sigma_{\rm{DRW}}$, $\tau_{\rm{DRW}}$, and $\sigma_{\rm{n}}$. We use Markov chain Monte Carlo (MCMC) implemented in the \textsc{emcee} package\cite{emcee} to sample the joint posterior probability density, with uniform priors for all parameters. We take the 16th and 84th percentiles of the marginalized posterior distributions for each parameter to estimate the $1\sigma$ uncertainties. To ensure the chains have sufficiently converged, we check that the auto-correlation function of the $\chi^2$ (model$-$data) residuals are consistent with white noise. We also performed model fitting using a generalization of Equation~(\ref{eq:drw}) to a higher-order CARMA model [e.g.~\cite{kelly14}, their equation 1] that allows more features in the PSD. We found in each case that the ACFs of the $\chi^2$ residuals were consistent within $3\sigma$ with Gaussian white noise for both the DRW and higher-order CARMA models, and the PSDs from both models overlap within 1$\sigma$. The quality of the available light curves does not justify more complicated models beyond the DRW model. Even if a higher-order CARMA model provides a better fit to the data, it is more difficult to physically interpret the multiple features (characteristic timescales) associated with those models.

\begin{figure*}[!t]
\centering
\renewcommand\thefigure{S1}
\includegraphics[width=1.0\textwidth]{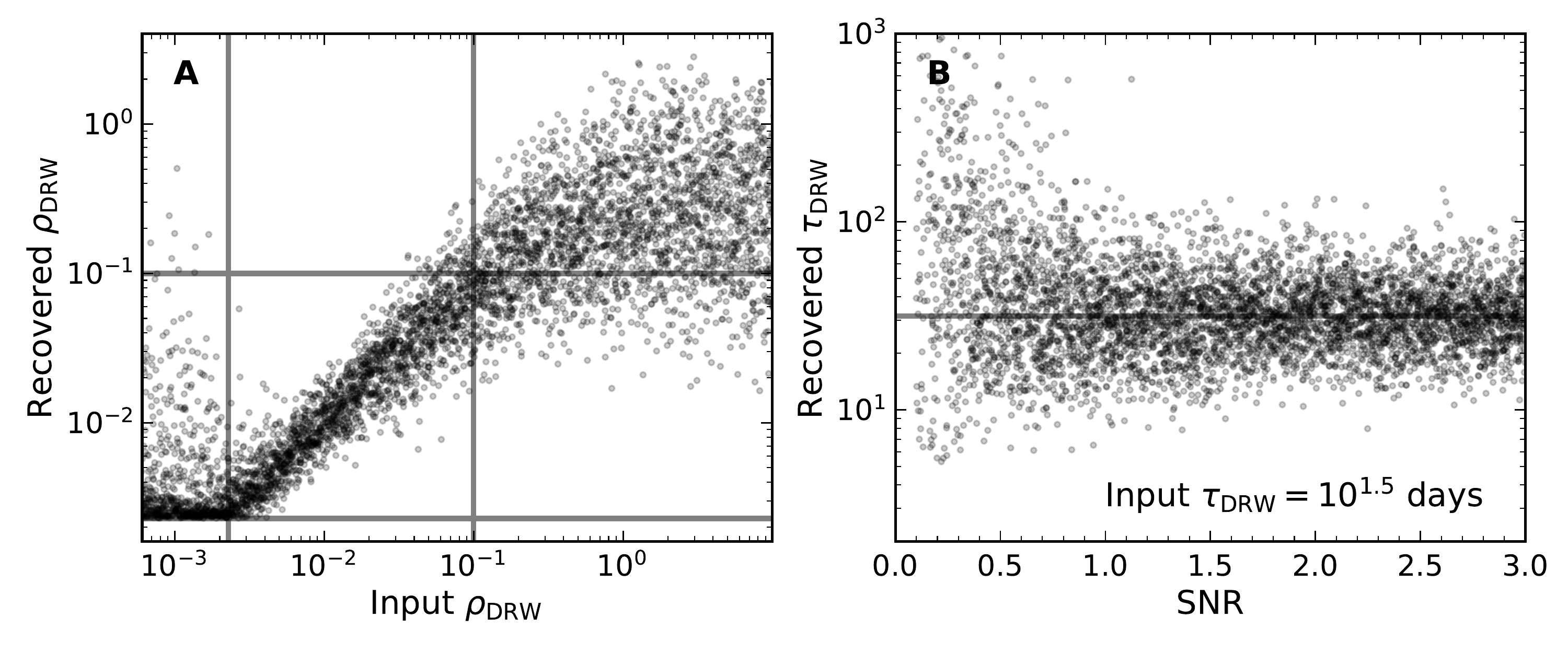}
\caption{\textbf{Recovered $\tau_{\rm{DRW}}$ for simulated light curves.} The recoverability of $\rho_{\rm{DRW}} = \tau_{\rm{DRW}}/\rm{baseline}$ is plotted for varying input $\rho_{\rm{DRW}}$ at constant SNR (A) and varying SNR at constant $\rho_{\rm{DRW}}$ (B). The pair of (vertical or horizontal) grey lines in panel A correspond to the cadence and $0.1 \times$ the light curve length. The horizontal grey line in panel B corresponds to the input $\tau_{\rm{DRW}}$. Input light curves are 6 years in length with typical 100-day seasonal gaps and a 5 day cadence. Better sampling with smaller gaps in the cadence improves the recoverability of the input parameters.} \label{fig:duration}
\end{figure*}

The recoverability of the damping timescale depends on the duration and signal-to-noise ratio (SNR) of the light curve. Previous work demonstrated that light curves of insufficient length (shorter than about 10 times the true value of $\tau_{\rm{DRW}}$) lead to a biased posterior probability distribution for $\tau_{\rm{DRW}}$, which typically saturates at 20 -- 30\% of the light curve length\cite{koz17}. We verify this behavior using simulated DRW light curves to test the recoverability of varying input $\tau_{\rm{DRW}}$. The simulations are similar to earlier work\cite{kelly09,koz10}, and the code to generate stochastic light curves is included in the python notebook at \url{https://github.com/burke86/taufit/tree/master/paper}. We find that when the true value of $\tau_{\rm{DRW}}$ is less than the cadence, the recovered $\tau_{\rm{DRW}}$ is close to the typical cadence of the light curve (regardless of larger seasonal gaps). If the uncertainties are larger than the amplitude of variability $\sigma_{\rm{DRW}}$, in the case of low signal-to-noise variability, the recovered $\tau_{\rm{DRW}}$ is larger than the true value. Figure~\ref{fig:duration} shows these effects in our simulations. We also require that the variability is AGN-like (i.e., red noise) by rejecting noisy light curves which are dominated by white noise. We compute the ACF of each light curve and reject those light curves that have ACF consistent with Gaussian white noise within the predicted $3\sigma$ confidence bands. Our final criteria to select a subsample with reliable measurements of the damping timescale are:

\begin{enumerate}
    \item Baseline: $\tau_{\rm{DRW}} < 0.1 \times \text{baseline}$\ ,
    \item Sampling: $\tau_{\rm{DRW}} >  \text{cadence}$\ ,
    \item Signal-to-noise: $\sigma_{\rm{DRW}}^2 > \sigma_{\rm{n}}^2 + \overline{dy}^2$\ ,
    \item AGN-like variability: ACF inconsistent ($3\sigma$) with white noise,
\end{enumerate}
where baseline is the light curve length, cadence is the mean cadence, and $\overline{dy}$ is the mean size of the quoted light curve uncertainties. 

\begin{figure*}[!t]
\centering
\renewcommand\thefigure{S2}
\includegraphics[width=0.8\textwidth]{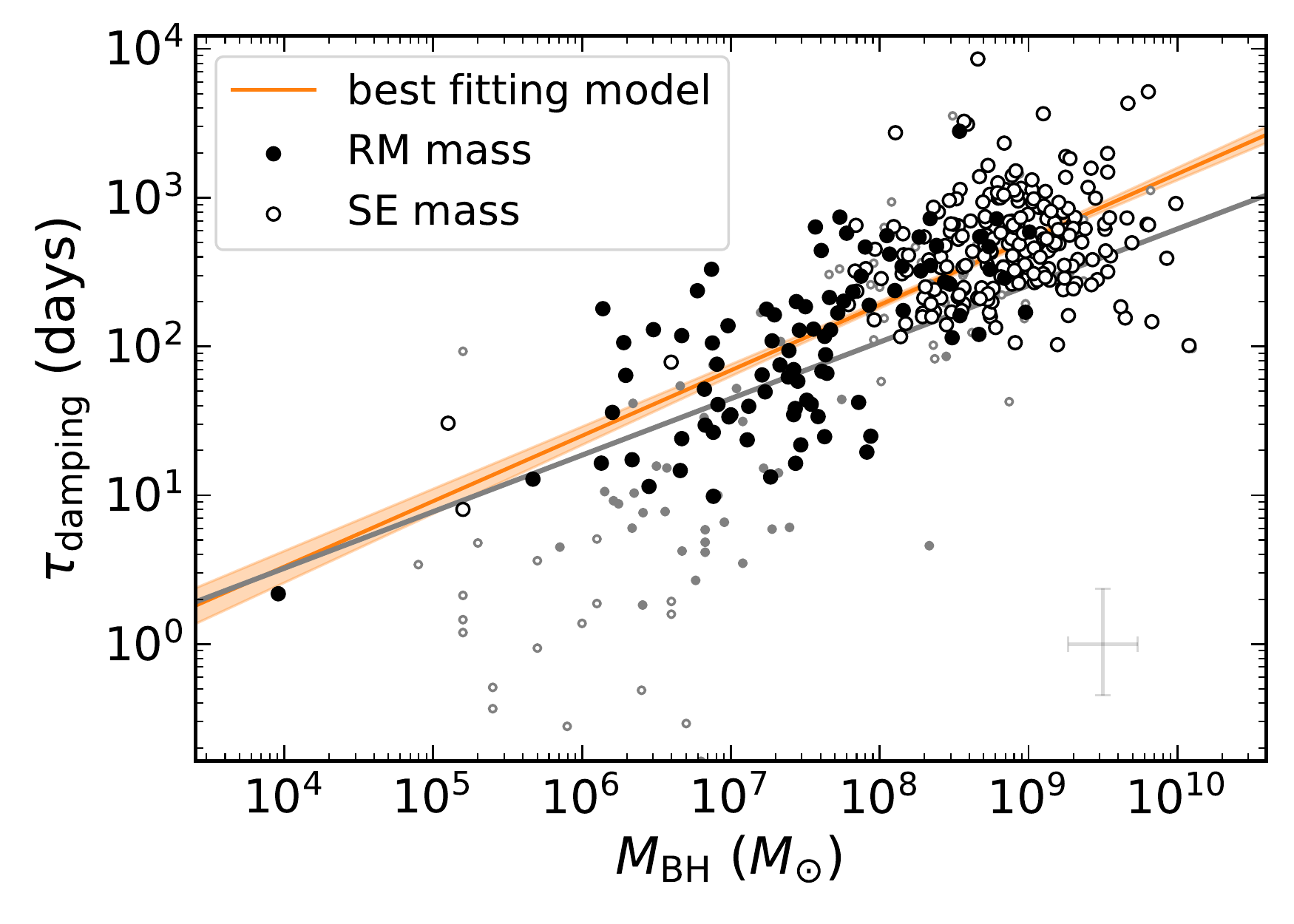}
\caption{\textbf{Optical variability damping timescale as a function of SMBH mass for the initial sample.} $\tau_{\rm{damping}}$ is plotted as a function of SMBH mass for all $\sim$ 400 AGNs in our initial sample. The large black circles are light curves of any length which satisfy the SNR, cadence, and ACF quality requirements. The small grey points are the remaining light curves of poor quality, regardless of SNR, cadence, or ACF (most of the extreme outliers). The typical $1\sigma$ uncertainty is shown with the error bar at the bottom-right. The best-fitting linear model from the final sample (as in Fig.~1) is shown as the grey line. The orange line and shaded band are the best fitting model and $1\sigma$ uncertainty for the large data points regardless of duration. This best-fitting relation is $\tau_{\rm{damping}} = 199^{+11}_{-11}\ {\rm{days}}\  \left(\frac{M_{\rm{BH}}}{10^8\ M_{\odot}}\right)^{0.44^{+0.02}_{-0.02}}$ with an intrinsic $1\sigma$ scatter of $0.35^{+0.02}_{-0.02}$ dex. The correlation is similar for the initial and final samples, but the initial sample has larger scatter. } \label{fig:tau_mass_all}
\end{figure*}

Figure 1 shows our results for the final sample of $\sim 60$ AGNs that pass our selection criteria. Figure S2 shows the results for the larger initial sample. As expected, the scatter in Figure~\ref{fig:tau_mass_all} is substantially larger than that in Figure~\ref{fig:tau_mass} due to low-quality measurements from light curves that do not pass our criteria. Most of the systems in our initial sample only slightly missed the duration requirement, hence the measured $\tau_{\rm damping}$ is not severely biased on average. We find that the correlation between SMBH mass and damping timescale also appears in this larger but lower quality sample. We find a mass slope closer to 0.5 than in Figure 1. This best-fitting relation is $\tau_{\rm{damping}} = 199^{+11}_{-11}\ {\rm{days}}\  \left(\frac{M_{\rm{BH}}}{10^8\ M_{\odot}}\right)^{0.44^{+0.02}_{-0.02}}$ with an intrinsic $1\sigma$ scatter of $0.35^{+0.02}_{-0.02}$ dex. However, the association of the damping timescale with the thermal timescale of the UV disk remains.

Our timing analysis has fitted the light curves with Gaussian process models in the time domain. Traditional PSD analysis performed in the frequency domain is much more demanding on the quality of the time series data. For our AGN sample, the optical light curves are irregularly sampled (except for the TESS light curve for NGC 4395), and there are seasonal gaps in multi-year light curves. These windowing effects severely impact PSD analysis in the frequency domain, making it much more challenging to recover the damping timescale than fitting with CARMA models\cite{kelly09,Burke20}. To demonstrate this point, we generated random light curves using DRW model parameters for our final sample, with the same cadence, duration and signal-to-noise ratio as the real data, and fitted DRW models to the simulated light curves in the time domain and used PSD analysis to retrieve the input break timescale. 

For the PSD analysis, we use the Lomb--Scargle periodogram\cite{Lomb76,Scargle82}, with PSD uncertainties estimated using the bootstrap technique. We resample each light curve and take the 16th and 84th percentiles to estimate the $1\sigma$ uncertainties of the measured PSD. Then, we bin the PSD (median and errors) in equal $\log_{10} f$ spacing. The resulting binned PSD is fitted with a broken power law model of the form $P \propto 1/[(f/f_{\rm{br}})^{a} + (f/f_{\rm{br}})^b]$ using the Levenberg--Marquardt method for nonlinear least-squares minimization, where $P$ is the PSD amplitude, $f_{\rm br}$ is the break frequency, and $a$ ($b$) is the slope of the power law at the high-(low-)frequency end. We discard fits which are unable to constrain $f_{\rm{br}}$ due to PSD noise and artifacts. The fitted PSD slopes can deviate from the DRW model due to well-known effects such as red noise leakage, sampling and windowing effects\cite{Uttley02}. 

\begin{figure*}[!t]
\centering
\renewcommand\thefigure{S3}
\includegraphics[width=0.85\textwidth]{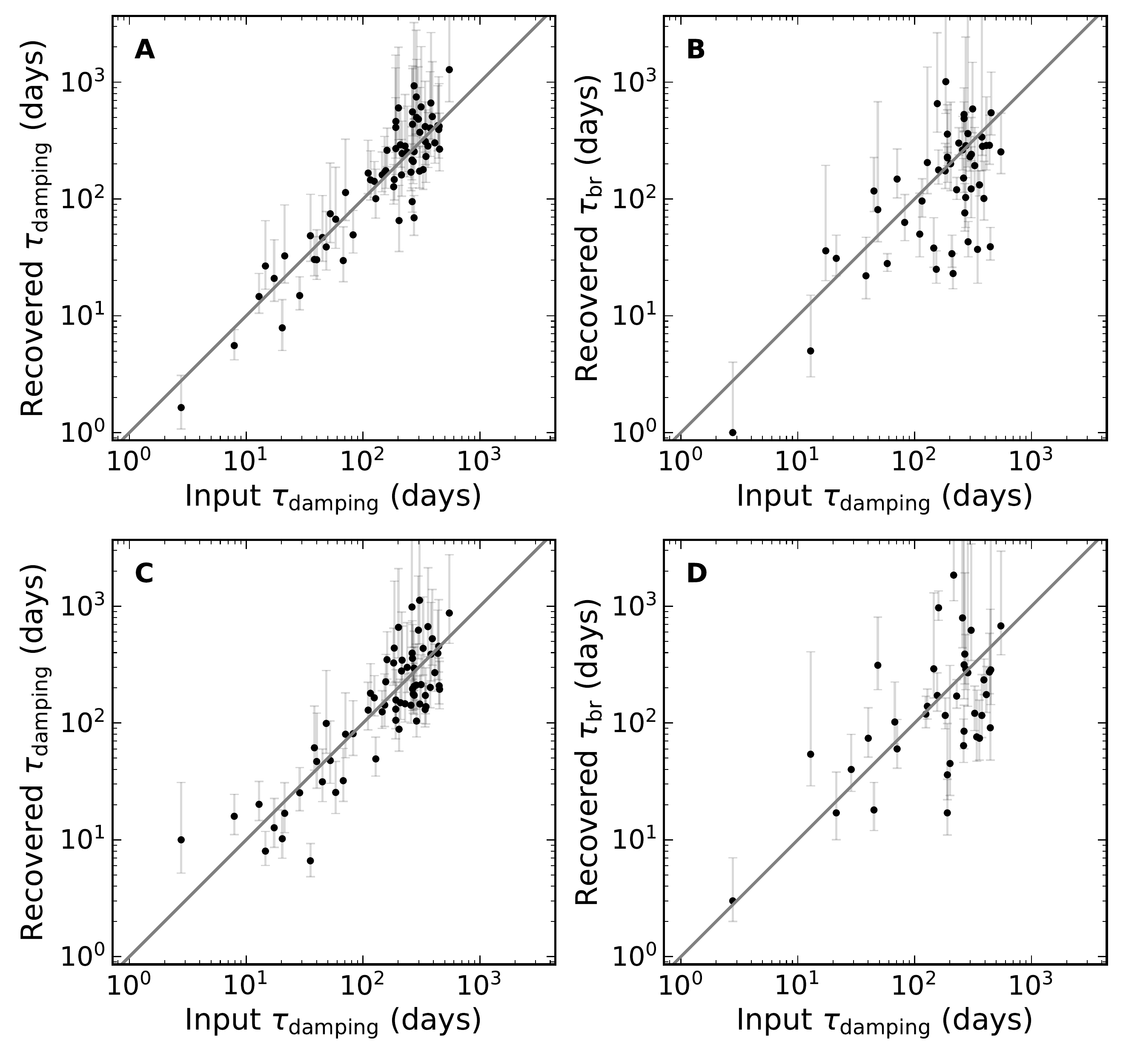}
\caption{\textbf{Recovered $\tau_{\rm{DRW}}$ and $\tau_{\rm{br}}$ for simulated light curves.} The recoverability of $\tau_{\rm{DRW}}$ from fitting DRW models and $\tau_{\rm{br}}$ from PSD analysis are shown for simulated light curves. Panels A and B are the results from DRW light curve simulations, and panels C and D are the results from forced DRW model fits to a non-DRW input light curve with a low-frequency PSD slope of $-1$ \cite{Timmer_Kong95}. Forced DRW fits to non-DRW input light curves can still recover the damping timescale, albeit with larger dispersion. Simulated input light curves are generated with the same cadence/sampling, length, flux uncertainty, and DRW parameters as the real light curves in our final sample of $\sim$ 60 AGNs. The PSD analysis cannot recover the input break timescale to the same precision as the DRW fits. All error bars are $1\sigma$. } \label{fig:tauintauout}
\end{figure*}

Figure~\ref{fig:tauintauout} compares the recoverability of the input break timescale using the DRW method and the periodogram analysis. As we expected, fitting the light curve in the time domain can recover the input break timescale more accurately than periodogram analysis, given the typical quality of the light curves. The periodogram analysis often cannot recover the correct PSD form, due to windowing effects such as noise leakage and aliasing, which is a worse problem for multi-season light curves. We also find that even if the low-frequency slope differs from a DRW with a slope of $-1$, similar to what is observed in some X-ray PSDs\cite{McHardy04}, forced fitting of a DRW model can still recover the correct input break timescale (Fig.~\ref{fig:tauintauout}C).

\begin{figure*}[!t]
\centering
\renewcommand\thefigure{S4}
\includegraphics[width=0.65\textwidth]{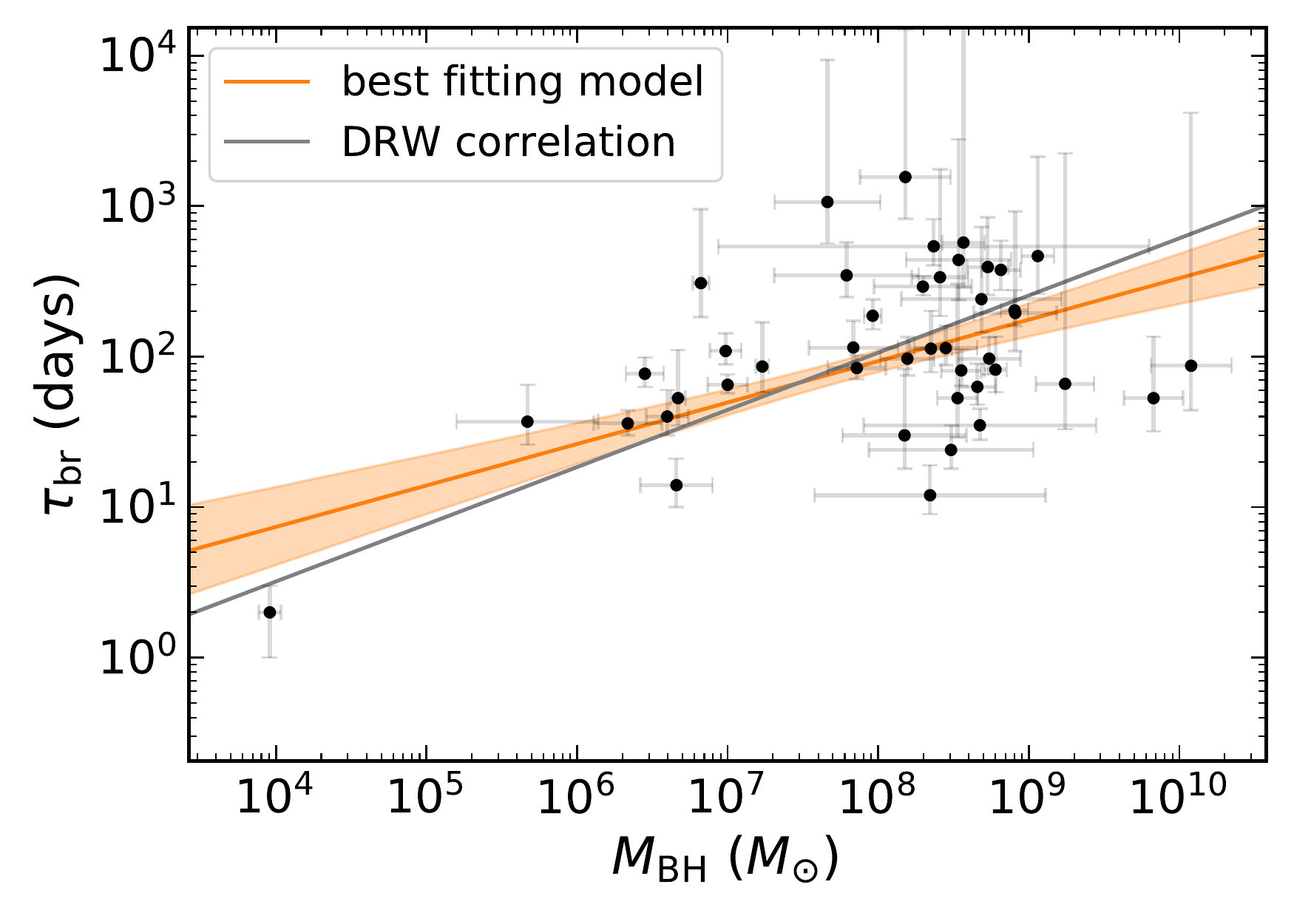}
\caption{\textbf{Optical variability from the PSD analysis in the frequency domain.} The black circles with $1\sigma$ error bars are the measured break timescales for our final sample from the PSD analysis. There is a correlation between the rest-frame break timescale $\tau_{\rm{br}}$ measured from a broken power law fit to AGN PSDs and SMBH mass. The orange line and shaded band are the best fitting model and $1\sigma$ uncertainty. The grey line is the best-fitting $\tau_{\rm{damping}}-M_{\rm{BH}}$ relation from Fig. 1. The two methods are broadly consistent. } \label{fig:taubr_mass}
\end{figure*}

We also performed PSD analysis for the real light curves in our final high-quality sample, and present the results in Figure~\ref{fig:taubr_mass}. The scatter is substantially larger, but a correlation is still present, which is roughly consistent (within $2\sigma$) with our fiducial result based on the DRW method. The PSD analysis shows a flattening of the PSD towards the low-frequency end in most objects, although the location of the break frequency/timescale often cannot be accurately determined, especially for the multi-year light curves with seasonal gaps. 

Even if some AGN light curves in our sample deviate from the DRW model, forced fitting of a DRW model provides a sufficient approximation, because the resulting damping timescale correlates with the SMBH mass. We were unable to identify any selection effects or modeling systematics that can artificially induce such a correlation. 

\begin{figure*}[!t]
\centering
\renewcommand\thefigure{S5}
\includegraphics[width=0.85\textwidth]{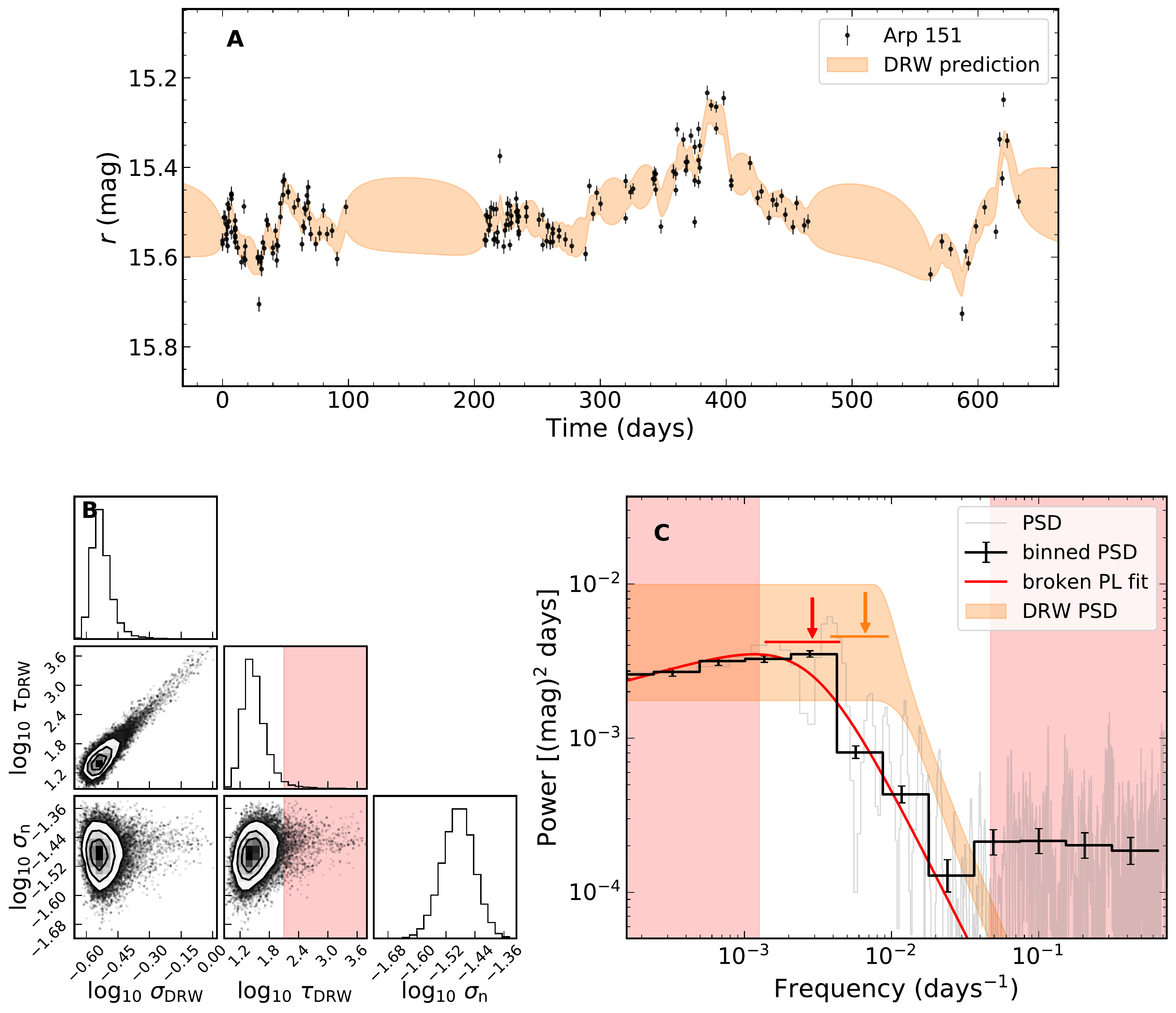}
\caption{\textbf{Example DRW modeling of Arp 151.} Panel A shows the $r$-band light curve of Arp 151 ($M_{\rm{BH}} = 10^{6.67\pm0.05} M_{\odot}$) and the best-fitting DRW model with 1$\sigma$ uncertainty (orange shaded area). Panel B shows the posterior probability distributions for the fitted DRW parameters and their covariance. In the covariance panels, the contours trace the 1, 2, 3$\sigma$ levels overplotted on the sample density map (black being higher density) with individual samples in the lowest-density regions shown as black points. Panel C shows the normalized PSD and binned PSD with 1$\sigma$ uncertainties. The best-fitting broken power-law model is shown as a red line. The 1$\sigma$ range of the DRW PSD from the posterior prediction is the orange shaded area. The corresponding break frequency $f_{\rm{br}}$ (from the broken power law fit) and $1 / (2 \pi \tau_{\rm{DRW}})$ (from the DRW fitting) are shown as equivalently colored arrows with the line segment below indicating the 1$\sigma$ unceratinty. The red shaded regions correspond to periods greater than 20\% the light curve length (in panels B and C) and less than the mean cadence (panel C), where the PSD is not well sampled. } \label{fig:example}
\end{figure*}

We present the timing analysis for all AGNs in our sample. An example light curve, DRW model, and PSD analysis is shown in Figure~\ref{fig:example}. The full figure set for the initial sample is available at \cite{Zenodo}.

We focus on the physical dependences of the damping timescale. The asymptotic variability amplitude $\sigma_{\rm DRW}$ in the DRW model has been demonstrated to correlate with wavelength, luminosity and SMBH mass of the AGN\cite{kelly09,MacLeod10}. We do not examine the wavelength dependence of the damping timescales directly, because most of our light curves from reverberation mapping samples cover only one band. The wavelength dependence of the damping timescale has been taken from previous work\cite{MacLeod10}.

\subsection*{Regression Analysis}

We perform linear regression between two physical quantities incorporating measurement uncertainties of both quantities. These measurement uncertainties are slightly asymmetric in general, therefore we symmetrize them by taking the mean. We use a hierarchical Bayesian model for fitting a line to data with measurement errors\cite{Kelly07}.

\begin{figure*}[ht]
\centering
\renewcommand\thefigure{S6}
\includegraphics[width=0.95\textwidth]{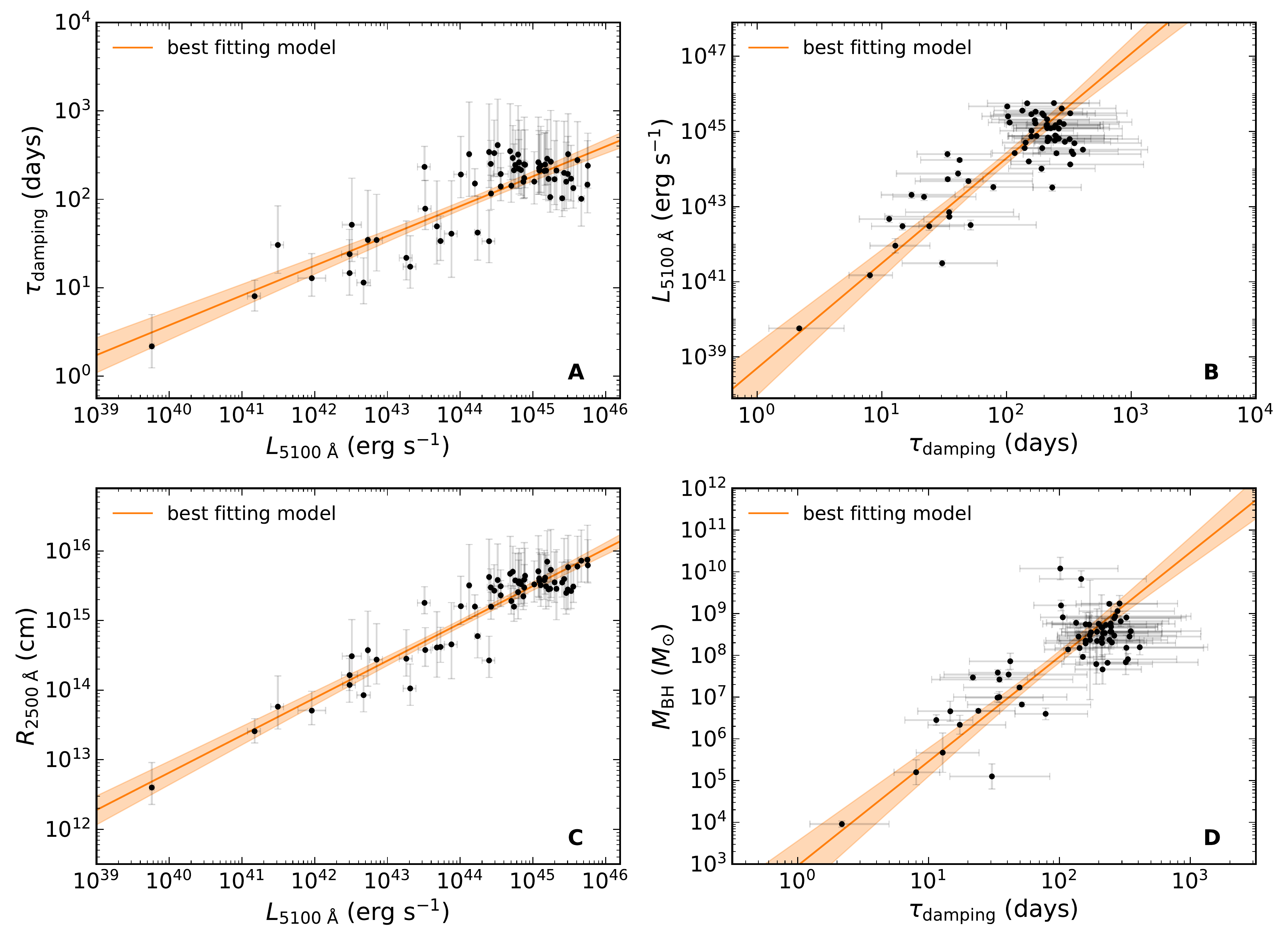}
\caption{\textbf{Correlations in other parameters.} Panels ABCD show the correlations among AGN luminosity $L_{5100\,\textrm{\AA}}$, UV-emitting accretion disk size $R_{\rm 2500\,\textrm{\AA}}$ and the damping timescale $\tau_{\rm damping}$. The characteristic emission radius is computed as $R \propto M_{\rm BH}^{1/3}\tau_{\rm{damping}}^{2/3}$, scaled to rest-frame $2500$ \AA\ according to the observed wavelength dependence of the damping timescale\cite{MacLeod10}, assuming $\tau_{\rm damping}$ is the local thermal timescale with $\alpha=0.05$. We overplot the best-fitting model relation (orange line) and $1\sigma$ uncertainty (orange shaded area) in each panel. All error bars are $1\sigma$. } \label{fig:regressions}
\end{figure*}

In addition to the regression fits shown in Figure~\ref{fig:tau_mass} and Figure~\ref{fig:radius_mass}, we show additional correlation analyses in Figure~\ref{fig:regressions}. These best-fitting model relations and $1\sigma$ intrinsic scatter (shown in the parentheses at the end) are:

\begin{equation}
    \tau_{\rm{damping}} = 10^{1.92^{+0.05}_{-0.05}}\ {\rm{days}}\ \left(\frac{L_{5100\ \textrm{\AA}}}{10^{44}\ {\rm erg\ s}^{-1}}\ \right)^{0.33^{+0.04}_{-0.04}}\ (0.11)\ , \tag{S5}
\end{equation}

\begin{equation}
    L_{5100\ \textrm{\AA}} = 10^{44.31^{+0.15}_{-0.15}}\ {\rm{erg\ s}^{-1}}\ \left(\tau_{\rm{damping}} / 100\ \rm{days} \right)^{2.83^{+0.36}_{-0.37}}\ (0.40)\ , \tag{S6}
\end{equation}

\begin{equation}
    R_{2500\ \textrm{\AA}} = 10^{14.95^{+0.05}_{-0.05}}\ {\rm{cm}}\ \left(\frac{L_{5100\ \textrm{\AA}}}{10^{44}\ {\rm erg\ s}^{-1}}\ \right)^{0.53^{+0.04}_{-0.04}}\ (0.10)\ ,
    \tag{S7}
\end{equation}

\begin{equation}
    M_{\rm{BH}} = 10^{7.97^{+0.14}_{-0.14}} M_{\odot}\ \left(\tau_{\rm{damping}} / 100\ \rm{days} \right)^{2.54^{+0.34}_{-0.35}}\ (0.33)\ . \tag{S8}
\end{equation}

In all these correlations, we have verified that the residuals do not depend on the Eddington ratio, which could be due to the limited range of $L_{\rm bol}/L_{\rm Edd}$ probed by our sample. The correlations between AGN luminosity/SMBH mass and observed (rest-frame) damping timescale provide empirical estimators for luminosity and SMBH mass, and the $1\sigma$ intrinsic scatter is $\sim 0.3-0.4$\,dex in both cases. 

As in the radius-mass relation, there is a correlation between the disk radius and the continuum luminosity $R_{\rm 2500\,\textrm{\AA}}\propto L^{0.53}$. Under the standard accretion disk model, the continuum luminosity at a given rest-frame wavelength can be derived by integrating over different radii\cite{Davis_Laor} such that $L_{\nu}\propto \nu^{1/3}\dot{M}_{\rm BH}^{2/3}M_{\rm BH}^{2/3}\cos i$, where $i$ is the inclination angle of the disk relative to the line-of-sight. Again if we assume constant Eddington ratio and radiative efficiency, we have $\dot{M}_{\rm BH}\propto M_{\rm BH}$ and $M_{\rm BH}\propto L_{\nu}^{3/4}$. Because we have an observed relation $R_{\rm 2500\,\textrm{\AA}}\propto M_{\rm BH}^{2/3}$ we expect a disk radius-luminosity relation $R_{\rm 2500\,\textrm{\AA}}\propto L_{\nu}^{0.5}$. The observed relation shown in Figure~\ref{fig:regressions} has a slope of $0.53\pm 0.04$. 

We have visually inspected the timing analysis results for outliers from the average correlations, in particular the two most massive SMBHs shown in Fig.~\ref{fig:tau_mass}. The damping timescale appears to be well constrained in these two systems. However, because there might be other processes that affect the shape of the PSD at different timescales, it is possible that the damping timescale we measure in the two most massive SMBHs is not the correct characteristic variability timescale that we are searching for. Nevertheless, these two objects are still consistent with the average relation given the large uncertainties in both axes. Excluding these two objects in the regression does not substantially change the results. 

Our adopted SMBH mass measurement\cite{Woo} of NGC 4395 differs from previous measurements by an order of magnitude\cite{Peterson_4395,denbrok}. If we instead adopt a SMBH mass\cite{Peterson_4395} of $3.6\times10^5\,M_\odot$ for NGC 4395, we derive a nearly identical correlation between the damping timescale and BH mass: $\tau_{\rm{damping}} = 105^{+11}_{-12}\ {\rm{days}}\  \left(\frac{M_{\rm{BH}}}{10^8\ M_{\odot}}\right)^{0.39^{+0.05}_{-0.05}}$. If we exclude NGC 4395 entirely from the sample, we also derive a nearly identical correlation: $\tau_{\rm{damping}} = 106^{+11}_{-13}\ {\rm{days}}\  \left(\frac{M_{\rm{BH}}}{10^8\ M_{\odot}}\right)^{0.38^{+0.05}_{-0.05}}$.

\subsection*{Unifying accretion variability timescales}\label{sec:disc}

We extend our timing analysis to other accreting systems using published measurements \cite{Scaringi15}, which include optical variability for accreting white dwarfs (WDs) of the nova-like class and X-ray variability for stellar-mass black holes and AGNs. Because the X-ray AGNs and WDs may accrete at very different $L/L_{\rm Edd}$ from our optical AGNs, we scale their measured variability timescales to the mean $L/L_{\rm Edd}=0.15$ of our AGN sample using the published relation \cite{Scaringi15}. Figure~\ref{fig:tau_mass} shows the variability timescale as a function of accretor mass, and the locations of the WDs are consistent with the extrapolation from AGNs assuming a mass slope of 0.5 expected from the standard accretion disk model. Accretion disks in binary stars may follow a similar relation\cite{Rucinski18}.

\begin{figure*}[t]
\centering
\renewcommand\thefigure{S7}
\includegraphics[width=0.8\textwidth]{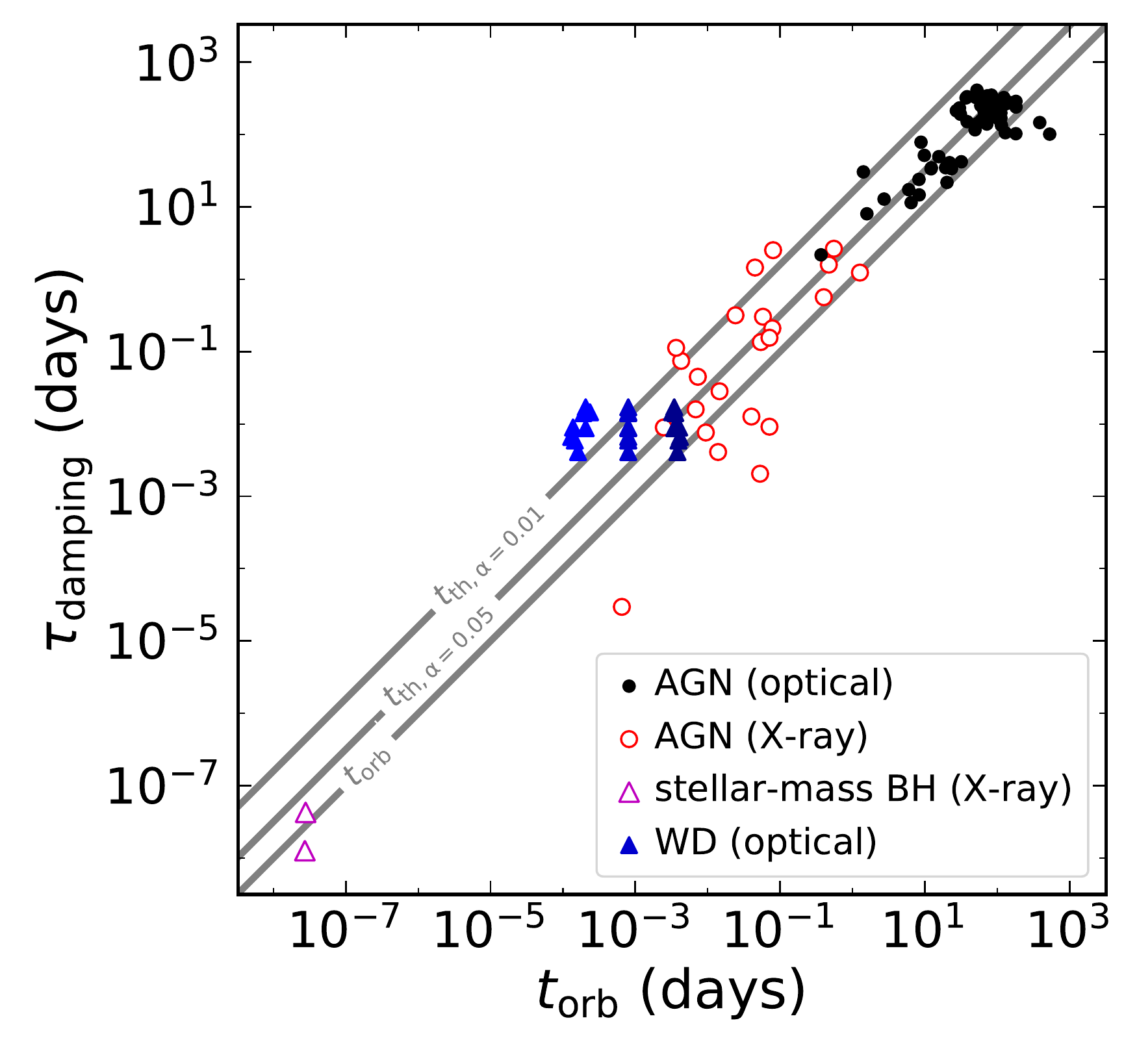}
\caption{\textbf{Variability timescales as a function of the orbital time.} The variability timescale is measured for different types of accreting systems and at both optical and X-ray wavelengths (black filled circles: AGNs at optical wavelengths; red open circles: AGNs at X-rays; filled triangles: WDs; open triangles: stellar-mass black holes). The thermal and orbital timescales for optical AGNs are calculated at the rest-frame 2500\,\AA\ emitting radius, estimated using the accretion disk sizes measured from microlensing\cite{Morgan18}. The timescales for X-ray AGNs and stellar-mass black holes are calculated at the ISCO radius ($3R_{S}$). The timescales for nova-like accreting white dwarfs are calculated at the emitting radius estimated with three different methods (Materials and Methods). The optical variability timescales for the accretion disk are consistent with the thermal timescale, with $\alpha=0.05$ for AGNs and $\alpha=0.01-0.05$ for WDs. The X-ray variability timescales are consistent with either the orbital timescale or the thermal timescale, given the large scatter.} \label{fig:tau_orb}
\end{figure*}

We further compare the measured variability timescale with the orbital or thermal timescale at the effective emitting radius in Figure~\ref{fig:tau_orb}. To estimate the effective 2500\,\AA\ radius for AGN accretion disks, we use the disk size--mass relation at rest-frame 2500\,\AA\ measured from microlensing\cite{Morgan18}. For the effective emitting radius of WDs, we use three different estimates to enclose the possible range: (i) extrapolation from the disk size--mass relation measured from microlensing of AGNs; (ii) the theoretical radius from the standard accretion disk theory [e.g. \cite{Morgan10}, their equation 2]; (iii) the WD surface radius compiled in previous work \cite{Scaringi15}. For the X-ray emitting radius in stellar-mass black holes and AGNs, we adopt the published estimates \cite{Scaringi15}, which are the ISCO radii ($3R_S$) assuming a BH spin parameter of 0.8. Despite the large range of timescales probed by these different accretors, the measured characteristic variability timescale lies close to the local orbital timescale for X-ray corona emission, and the local thermal timescale for optical accretion disk emission. For the X-ray timescales this agreement is within a factor of $\sim 5$ ($1\sigma$) of the orbital time, given the large dispersion in individual X-ray AGN and the small number of stellar-mass black holes. 

Because the emitting radius of nova-like accreting WDs likely extends close to the WD surface\cite{Scaringi15}, Fig.~\ref{fig:tau_orb} implies $\alpha\sim 0.01$, smaller than $\alpha\sim 0.05$ for AGN accretion disks from our analysis. We interpret this as due to nova-like WD accretion disks being gas-pressure-dominated and lacking the high opacities that might drive convection, which would be consistent with magnetorotational instability (MRI) simulations that predict $\alpha\sim 0.01$ \cite{MRI,Hawley13}. The association of the optical variability damping timescale with the local thermal timescale therefore seems the same for both AGN and accreting WD accretion disks despite different disk structures. 

\section*{Supplementary Text}

\subsection*{Comparison with earlier observations}

Previous observational studies examined the potential correlation between a characteristic optical variability timescale with the SMBH mass of the AGN. A structure function analysis of 13 AGNs in the UV/optical found evidence of increasing characteristic timescale with SMBH mass \cite{Collier_Peterson01}. However, the statistics were poor and the correlation was not well constrained. Previous work using DRW model fitting to optical light curves of 70 AGNs found a correlation between the damping timescale and the SMBH mass with a slope ($0.56\pm0.14$) \cite{kelly09} consistent with our results ($0.38\pm0.04$) within $\sim 1\sigma$, but the uncertainties of the slope and normalization are substantially larger than ours. However, most of the light curves in their sample do not pass the duration criterion, and the inferred damping timescale is likely biased to some extent\cite{koz17}. Similarly, DRW model fitting to $\sim 9000$ quasars in the SDSS Stripe 82 region\cite{MacLeod10} revealed a correlation between the damping timescale and the SMBH mass with a slope much shallower ($0.21\pm0.07$) than ours. Many of the light curves in this sample also do not pass our duration criterion and the individual damping timescales have large uncertainties. Another study \cite{Simm16} of optical light curves of a large quasar sample with multi-year light curves found no correlation between the break timescale and the SMBH mass of the quasar. 

Those previous studies were limited by the dynamic range in mass. Measurements of individual damping timescales have large uncertainties, given the typical quality of these light curves and the difficulty to precisely measure the curvature of the PSD. Vetting of the light curves is necessary to ensure the robustness of the measured damping timescale, especially requiring sufficiently long duration of the light curve to constrain the damping of variability on long timescales. Our sample spans the entire mass range of SMBHs and has sufficiently long optical light curves to robustly measure the damping timescale\cite{methods}. The longest duration of our light curves is $\sim 20$\,years with hundreds of epochs, sufficient to constrain the damping timescale in the $\sim 100-300$\ day range. The slope and normalization of this correlation are both necessary for interpretation of the nature of AGN optical variability and accretion models. For example, although \cite{kelly09} reported a similar correlation, the uncertainty in the normalization is more than a factor of ten, too large to distinguish between the thermal and orbital/dynamical timescales.  \\

\subsection*{Theoretical implications }

MHD turbulence in hot, ionized accretion disks can drive variability at all radii in the disk, with variations in the turbulent heating rate resulting in luminosity variations at or longer than the local thermal time of the accretion disk.  In standard accretion disk models, different photon wavelengths originate from different radii in the disk, and one might then
expect that the luminosity at different wavelengths would vary with different characteristic time scales that match their local thermal times. This is not what is observed in AGNs\cite{MacLeod10,Suberlak21}, instead, only the UV emission varies with characteristic timescales equal to the thermal time
(assuming $\alpha=0.05$). The timescales at optical wavelengths are nearly
the same as the ultraviolet thermal time, which is not consistent with the
thermal times at optically emitting radii.  

This suggests that the dominant variability originates within the disk at UV emitting radii, and that variability is then communicated in some way to other radii in the disk. We suggest several possible explanations for why variability might originate in the UV emitting radii.  These radii may reach interior temperatures near $10^5$~K where the Rosseland mean opacity of the material is enhanced by iron \cite{JIA20}. In the radiation pressure dominated conditions of an AGN accretion disk, this can drive convective motions that in turn interact with the MHD turbulence to drive surface density and luminosity fluctuations\cite{JIA20}. This convection can also enhance the viscosity parameter $\alpha$ to values in the range $0.05-0.1$, which would then be consistent with our chosen value of 0.05 needed to match the thermal time at UV emitting radii with our measured damping time\cite{JIA20,SCE18,HIR14}. (Simulations of MHD turbulence in accretion disks without imposing external vertical magnetic field find $\alpha\sim0.01$\cite{MRI,Hawley13}.) Radiation pressure from UV photons acting on spectral lines could also drive variable outflows from these radii, resulting in time-dependent mass loss\cite{PRO00}.

It is unclear what mechanisms might enable this variability to induce similar variability timescale at other radii. It has previously been suggested that different radii in an AGN accretion disk might be coupled together by large-scale magnetic fields, and variability at one radius might launch Alfv\'en waves into other radii to drive dissipative heating there at similar variability timescales to those at the launching radius\cite{Char}. A similar proposal has been suggested in nova-like white dwarf accretion disks:  these long-lived disks might build up large-scale magnetic fields that could enable mass exchange between different radii\cite{NIX19}.  Regardless of the physical mechanism, our results suggest that it is a ubiquitous process in accretion disks.

\clearpage

%Our final sample is restricted to the region \red{Indicate this region on the figure} where $\tau_{\rm{DRW}}$ is well-constrained. 

\newgeometry{margin=0.5cm}
\begin{landscape}
{\tiny
\begin{longtable}{lllrrrrrrrlllllrrrrrrrlrrrl}
\label{tab:1}
\\\caption{\textbf{Summary data for the final AGN sample.} We list the J2000 right ascension (RA), declination (DEC.), redshift ($z$), and light curve (LC) reference, along with other parameters. The ``Used Name'' column lists the source name in the original LC reference. SNR $= \sigma_{\rm{DRW}} / \sqrt{\sigma_{\rm{n}}^2 + \overline{dy}^2}$. If $L_{5100\ \textrm{\AA}}$ uncertainties are not given in the primary sources, we assume a fractional uncertainty of $20\%$ (typical of AGN variability). For a few objects, there are multiple light curves available, and we only include the $\tau_{\rm damping}$ measurement from the best light curve in the final sample. An equivalent table for the $\sim 400$ AGNs in the initial sample is provided in Data S1. The pivot wavelengths for different filters are 8500, 4371, 5478, 6504, 4702, 6175 \AA\ in $TBVRgr$ bands ($T$=TESS band).} \label{tab:data}\\
%\\\hline
\\\hline
Object Designation & Used Name & RA & DEC. & $z$ & $\log\ M_{\rm{BH}}$ & $\log\ L_{5100\ \textrm{\AA}}$ & band & $M_{\rm{BH}}$ Method & LC Ref. & $L_{5100\ \textrm{\AA}}$ Ref. & $M_{\rm{BH}}$ Ref. &  baseline & cadence & $\log \tau_{\rm{DRW,rest}}$ & SNR & ACF &  valid \\

{} & {} & [deg.] & [deg.] & & $\log\ \rm{[M_{\odot}}$] & $\log\ \rm{[erg\ s}^{-1}$] &  &  &  &  &  & [days] & [days] & $\log\ \rm{[days}$] & & &  & \\
\hline
%\endhead
%\endfoot
       NGC 4395 & NGC 4395 &  186.4536 &   33.5469 &  0.0011 &   3.96 $\pm$ 0.07 &  39.76 $\pm$ 0.03 &  TESS &  RM &      \cite{Burke20} &            \cite{Cho} &          \cite{Woo} &       27 &     0.02 &  0.3 $\pm$ 0.4 &  1.61 &      True &   True   \\
   MACHO J045614.18-673910.8 & 48.2620.2719 &   74.0591 &  -67.6530 &  0.2600 &   7.79 $\pm$ 0.48 &  44.01 $\pm$ 0.00 &     V &  SE &       \cite{Geha03} &      \cite{kelly09} &      \cite{kelly09} &     2619 &     7.25 &  2.3 $\pm$ 0.5 &  1.97 &      True &   True     \\
        Mrk 279 & Mrk 279 &  208.2643 &   69.3082 &  0.0300 &   7.54 $\pm$ 0.11 &  43.88 $\pm$ 0.00 &     R &  RM &   \cite{peterson04} &      \cite{kelly09} &      \cite{kelly09} &      527 &     4.25 &  1.6 $\pm$ 0.8 &  1.21 &      True &   True     \\
       NGC 5548 & NGC 5548 &  214.4981 &   25.1368 &  0.0170 &   7.82 $\pm$ 0.01 &  43.51 $\pm$ 0.00 &     R &  RM &   \cite{peterson04} &      \cite{kelly09} &      \cite{kelly09} &     4756 &     3.11 &  2.4 $\pm$ 0.2 &  3.75 &      True &   True     \\
       NGC 3783 & NGC 3783 &  174.7571 &  -37.7386 &  0.0090 &   7.47 $\pm$ 0.07 &  43.26 $\pm$ 0.00 &     R &  RM &   \cite{peterson04} &      \cite{kelly09} &      \cite{kelly09} &      226 &     3.22 &  1.3 $\pm$ 0.4 &  1.43 &      True &   True     \\
 DES J021822.51-043036.0 & DES J0218-0430 &   34.5938 &   -4.5100 &  0.8230 &   6.60 $\pm$ 0.14 &  43.52 $\pm$ 0.00 &     g &  SE &          \cite{Guo} &          \cite{Guo} &          \cite{Guo} &     2208 &    15.88 &  1.9 $\pm$ 0.3 &  1.45 &      True &   True     \\
        SDSS J024712.91-011106.4 & 2551748 &   41.8038 &   -1.1851 &  0.4863 &   8.18 $\pm$ 0.30 &  44.12 $\pm$ 0.01 &     g &  SE &         \cite{Yang} &       \cite{Shen11} &       \cite{Shen11} &     5543 &    32.80 &  2.5 $\pm$ 0.8 &  1.76 &      True &   True     \\
        SDSS J024422.20-011247.2 & 2570941 &   41.0925 &   -1.2131 &  1.6494 &   8.48 $\pm$ 0.55 &  45.21 $\pm$ 0.01 &     g &  SE &         \cite{Yang} &       \cite{Shen11} &       \cite{Shen11} &     7372 &    33.06 &  2.2 $\pm$ 0.3 &  3.12 &      True &   True     \\
        SDSS J024212.65-010339.6 & 2531595 &   40.5527 &   -1.0610 &  1.4368 &   8.57 $\pm$ 0.14 &  44.76 $\pm$ 0.03 &     g &  SE &         \cite{Yang} &       \cite{Shen11} &       \cite{Shen11} &     7458 &    34.05 &  2.4 $\pm$ 0.4 &  2.88 &      True &   True     \\
        SDSS J024259.02-001038.3 & 2562971 &   40.7459 &   -0.1773 &  0.7327 &   7.91 $\pm$ 0.13 &  44.47 $\pm$ 0.05 &     g &  SE &         \cite{Yang} &       \cite{Shen11} &       \cite{Shen11} &     7464 &    37.32 &  2.5 $\pm$ 0.4 &  1.87 &      True &   True     \\
        SDSS J025254.00+005832.2 & 2568243 &   43.2250 &    0.9756 &  1.2979 &   8.47 $\pm$ 0.09 &  44.81 $\pm$ 0.01 &     g &  SE &         \cite{Yang} &       \cite{Shen11} &       \cite{Shen11} &     7406 &    65.54 &  2.4 $\pm$ 0.6 &  3.46 &      True &   True     \\
        SDSS J024426.88-003028.4 & 2601455 &   41.1120 &   -0.5079 &  2.0831 &   9.19 $\pm$ 0.13 &  45.40 $\pm$ 0.04 &     g &  SE &         \cite{Yang} &       \cite{Shen11} &       \cite{Shen11} &     6616 &    38.92 &  2.0 $\pm$ 0.3 &  1.30 &      True &   True     \\
        SDSS J025007.03+002525.3 & 2590102 &   42.5293 &    0.4237 &  0.1978 &   7.96 $\pm$ 0.06 &  44.20 $\pm$ 0.00 &     g &  SE &         \cite{Yang} &       \cite{Shen11} &       \cite{Shen11} &     7464 &    27.24 &  2.2 $\pm$ 0.2 &  3.68 &      True &   True     \\
        SDSS J025401.56+000317.3 & 2479049 &   43.5065 &    0.0548 &  1.8317 &   8.91 $\pm$ 0.28 &  45.24 $\pm$ 0.02 &     g &  SE &         \cite{Yang} &       \cite{Shen11} &       \cite{Shen11} &     7458 &    35.18 &  2.0 $\pm$ 0.2 &  1.22 &      True &   True     \\
        SDSS J024959.78-000104.1 & 7910366 &   42.4991 &   -0.0178 &  1.5088 &   8.66 $\pm$ 0.12 &  45.09 $\pm$ 0.01 &     g &  SE &         \cite{Yang} &       \cite{Shen11} &       \cite{Shen11} &     7464 &    31.90 &  2.3 $\pm$ 0.4 &  2.35 &      True &   True     \\
        SDSS J024935.47+004144.5 & 2590128 &   42.3978 &    0.6957 &  1.9586 &   8.74 $\pm$ 0.28 &  45.02 $\pm$ 0.06 &     g &  SE &         \cite{Yang} &       \cite{Shen11} &       \cite{Shen11} &     7439 &    35.94 &  2.2 $\pm$ 0.3 &  2.45 &      True &   True     \\
        SDSS J024544.78-004415.4 & 2585077 &   41.4366 &   -0.7376 &  1.2250 &   8.69 $\pm$ 0.53 &  45.17 $\pm$ 0.00 &     g &  SE &         \cite{Yang} &       \cite{Shen11} &       \cite{Shen11} &     7372 &    38.80 &  2.4 $\pm$ 0.3 &  2.56 &      True &   True     \\
        SDSS J024547.59-000814.3 & 2585096 &   41.4483 &   -0.1373 &  1.6026 &   8.89 $\pm$ 0.15 &  45.08 $\pm$ 0.02 &     g &  SE &         \cite{Yang} &       \cite{Shen11} &       \cite{Shen11} &     7464 &    30.59 &  2.4 $\pm$ 0.6 &  2.65 &      True &   True     \\
        SDSS J024746.99-011334.3 & 2485925 &   41.9458 &   -1.2262 &  1.3853 &   8.76 $\pm$ 0.14 &  44.89 $\pm$ 0.01 &     g &  SE &         \cite{Yang} &       \cite{Shen11} &       \cite{Shen11} &     7372 &    37.23 &  2.4 $\pm$ 0.4 &  4.98 &      True &   True     \\
        SDSS J024204.58-003835.9 & 2498197 &   40.5191 &   -0.6433 &  2.2888 &   8.78 $\pm$ 0.07 &  45.56 $\pm$ 0.02 &     g &  SE &         \cite{Yang} &       \cite{Shen11} &       \cite{Shen11} &     7372 &    38.00 &  2.1 $\pm$ 0.3 &  1.81 &      True &   True     \\
        SDSS J024315.62-002032.3 & 2534406 &   40.8151 &   -0.3423 &  0.8059 &   8.56 $\pm$ 0.08 &  44.56 $\pm$ 0.10 &     g &  SE &         \cite{Yang} &       \cite{Shen11} &       \cite{Shen11} &     7458 &    29.25 &  2.3 $\pm$ 0.2 &  2.61 &      True &   True     \\
        SDSS J025221.84-003358.7 & 2593261 &   43.0910 &   -0.5663 &  2.0156 &   8.41 $\pm$ 0.19 &  45.18 $\pm$ 0.03 &     g &  SE &         \cite{Yang} &       \cite{Shen11} &       \cite{Shen11} &     7464 &    29.50 &  2.3 $\pm$ 0.5 &  2.00 &      True &   True     \\
        SDSS J025030.77-000801.7 & 2616632 &   42.6282 &   -0.1338 &  1.4602 &   9.23 $\pm$ 0.04 &  45.76 $\pm$ 0.00 &     g &  SE &         \cite{Yang} &       \cite{Shen11} &       \cite{Shen11} &     7464 &    40.34 &  2.4 $\pm$ 0.4 &  2.21 &      True &   True     \\
        SDSS J025131.66+003251.7 & 2580699 &   42.8819 &    0.5477 &  2.9191 &   8.74 $\pm$ 0.21 &  45.30 $\pm$ 0.03 &     g &  SE &         \cite{Yang} &       \cite{Shen11} &       \cite{Shen11} &     7433 &    33.63 &  2.2 $\pm$ 0.5 &  1.87 &      True &   True     \\
        SDSS J024511.93-011317.4 & 2567584 &   41.2997 &   -1.2215 &  2.4622 &   8.75 $\pm$ 0.18 &  45.43 $\pm$ 0.05 &     g &  SE &         \cite{Yang} &       \cite{Shen11} &       \cite{Shen11} &     7458 &    61.63 &  2.3 $\pm$ 0.8 &  1.59 &      True &   True     \\
        SDSS J025217.47-005249.4 & 2507583 &   43.0728 &   -0.8804 &  0.9113 &   8.58 $\pm$ 0.12 &  44.69 $\pm$ 0.01 &     g &  SE &         \cite{Yang} &       \cite{Shen11} &       \cite{Shen11} &     6721 &    32.31 &  2.5 $\pm$ 0.6 &  2.10 &      True &   True     \\
        SDSS J024442.77-004223.0 & 2598355 &   41.1782 &   -0.7064 &  0.6279 &   8.19 $\pm$ 0.18 &  44.52 $\pm$ 0.01 &     g &  SE &         \cite{Yang} &       \cite{Shen11} &       \cite{Shen11} &     7372 &    31.91 &  2.6 $\pm$ 0.6 &  3.36 &      True &   True     \\
        SDSS J024826.69-004144.5 & 2615997 &   42.1112 &   -0.6957 &  3.0023 &   9.83 $\pm$ 0.20 &  45.75 $\pm$ 0.02 &     g &  SE &         \cite{Yang} &       \cite{Shen11} &       \cite{Shen11} &     7372 &    32.05 &  2.2 $\pm$ 0.6 &  1.75 &      True &   True     \\
        SDSS J025311.71-004241.8 & 2509117 &   43.2988 &   -0.7116 &  1.5378 &   8.37 $\pm$ 0.34 &  45.11 $\pm$ 0.01 &     g &  SE &         \cite{Yang} &       \cite{Shen11} &       \cite{Shen11} &     7376 &    32.21 &  2.4 $\pm$ 0.5 &  2.15 &      True &   True     \\
        SDSS J025005.69-004054.8 & 2549476 &   42.5237 &   -0.6819 &  1.3170 &   8.53 $\pm$ 0.13 &  44.84 $\pm$ 0.01 &     g &  SE &         \cite{Yang} &       \cite{Shen11} &       \cite{Shen11} &     6649 &    32.12 &  2.3 $\pm$ 0.5 &  2.85 &      True &   True     \\
        SDSS J024823.52+003552.8 & 2579128 &   42.0980 &    0.5980 &  1.0149 &   8.29 $\pm$ 0.21 &  44.87 $\pm$ 0.01 &     g &  SE &         \cite{Yang} &       \cite{Shen11} &       \cite{Shen11} &     7458 &    32.15 &  2.2 $\pm$ 0.2 &  3.17 &      True &   True     \\
        SDSS J024512.12-011314.2 & 2593550 &   41.3005 &   -1.2206 &  2.4600 &   8.37 $\pm$ 1.43 &  45.53 $\pm$ 0.03 &     g &  SE &         \cite{Yang} &       \cite{Shen11} &       \cite{Shen11} &     7458 &    64.29 &  2.2 $\pm$ 0.4 &  1.38 &      True &   True     \\
        SDSS J025312.94-003729.6 & 2564901 &   43.3039 &   -0.6249 &  0.9821 &   8.45 $\pm$ 0.12 &  44.40 $\pm$ 0.01 &     g &  SE &         \cite{Yang} &       \cite{Shen11} &       \cite{Shen11} &     7458 &    30.32 &  2.5 $\pm$ 0.7 &  2.60 &      True &   True     \\
        SDSS J024257.22-004549.3 & 2605412 &   40.7384 &   -0.7637 &  1.7737 &   8.56 $\pm$ 0.36 &  44.88 $\pm$ 0.05 &     g &  SE &         \cite{Yang} &       \cite{Shen11} &       \cite{Shen11} &     7372 &    31.24 &  2.4 $\pm$ 0.7 &  3.10 &      True &   True     \\
        SDSS J024703.24-010032.0 & 2608769 &   41.7635 &   -1.0089 &  2.5314 &   8.35 $\pm$ 0.14 &  45.46 $\pm$ 0.03 &     g &  SE &         \cite{Yang} &       \cite{Shen11} &       \cite{Shen11} &     7464 &    31.90 &  2.2 $\pm$ 0.5 &  1.69 &      True &   True     \\
        SDSS J024531.54-002612.1 & 2555319 &   41.3814 &   -0.4367 &  2.0858 &   8.30 $\pm$ 0.32 &  45.32 $\pm$ 0.03 &     g &  SE &         \cite{Yang} &       \cite{Shen11} &       \cite{Shen11} &     7464 &    29.38 &  2.3 $\pm$ 0.7 &  1.55 &      True &   True     \\
        SDSS J024954.38+003654.4 & 2490551 &   42.4766 &    0.6151 &  1.7215 &   8.93 $\pm$ 0.11 &  45.25 $\pm$ 0.01 &     g &  SE &         \cite{Yang} &       \cite{Shen11} &       \cite{Shen11} &     7458 &    33.59 &  2.4 $\pm$ 0.7 &  3.16 &      True &   True     \\
        SDSS J025221.55+002832.5 & 2491276 &   43.0898 &    0.4757 &  1.2508 &   8.73 $\pm$ 0.14 &  45.09 $\pm$ 0.02 &     g &  SE &         \cite{Yang} &       \cite{Shen11} &       \cite{Shen11} &     6720 &    27.32 &  2.4 $\pm$ 0.5 &  2.00 &      True &   True     \\
        SDSS J024455.18-002501.6 & 2521752 &   41.2299 &   -0.4171 &  1.2998 &   9.24 $\pm$ 0.19 &  45.20 $\pm$ 0.01 &     g &  SE &         \cite{Yang} &       \cite{Shen11} &       \cite{Shen11} &     6721 &    21.40 &  2.5 $\pm$ 0.5 &  2.81 &      True &   True     \\
        SDSS J025329.35+002753.6 & 2495465 &   43.3723 &    0.4649 &  0.9552 &   8.81 $\pm$ 0.13 &  44.72 $\pm$ 0.01 &     g &  SE &         \cite{Yang} &       \cite{Shen11} &       \cite{Shen11} &     7433 &    31.23 &  2.5 $\pm$ 0.4 &  1.79 &      True &   True     \\
        SDSS J025151.53-000407.3 & 2500072 &   42.9647 &   -0.0687 &  2.0876 &   8.68 $\pm$ 0.77 &  45.16 $\pm$ 0.04 &     g &  SE &         \cite{Yang} &       \cite{Shen11} &       \cite{Shen11} &     7432 &    32.74 &  2.3 $\pm$ 0.5 &  2.17 &      True &   True     \\
        SDSS J025050.64+004503.2 & 2577933 &   42.7110 &    0.7509 &  2.0529 &   8.34 $\pm$ 0.77 &  45.48 $\pm$ 0.03 &     g &  SE &         \cite{Yang} &       \cite{Shen11} &       \cite{Shen11} &     7464 &    29.04 &  2.3 $\pm$ 0.5 &  1.56 &      True &   True     \\
        SDSS J024920.98+004206.5 & 2559705 &   42.3374 &    0.7018 &  1.5221 &   8.55 $\pm$ 0.13 &  44.88 $\pm$ 0.01 &     g &  SE &         \cite{Yang} &       \cite{Shen11} &       \cite{Shen11} &     7464 &    34.08 &  2.2 $\pm$ 0.3 &  3.22 &      True &   True     \\
        SDSS J024651.86-010732.5 & 2613947 &   41.7161 &   -1.1257 &  0.6218 &   8.14 $\pm$ 0.10 &  44.43 $\pm$ 0.02 &     g &  SE &         \cite{Yang} &       \cite{Shen11} &       \cite{Shen11} &     7458 &    61.13 &  2.1 $\pm$ 0.2 &  2.16 &      True &   True     \\
        SDSS J024347.38-010611.9 & 2607635 &   40.9474 &   -1.1033 &  3.9277 &  10.08 $\pm$ 0.27 &  45.67 $\pm$ 0.04 &     g &  SE &         \cite{Yang} &       \cite{Shen11} &       \cite{Shen11} &     7372 &    38.60 &  2.0 $\pm$ 0.5 &  1.54 &      True &   True     \\
        SDSS J025224.98+001308.0 & 2597127 &   43.1041 &    0.2189 &  1.2950 &   7.83 $\pm$ 0.30 &  44.80 $\pm$ 0.01 &     g &  SE &         \cite{Yang} &       \cite{Shen11} &       \cite{Shen11} &     7393 &    33.60 &  2.5 $\pm$ 0.7 &  3.91 &      True &   True     \\
        SDSS J024357.91-011330.7 & 2481079 &   40.9913 &   -1.2252 &  0.9037 &   8.45 $\pm$ 0.21 &  44.56 $\pm$ 0.01 &     g &  SE &         \cite{Yang} &       \cite{Shen11} &       \cite{Shen11} &     7458 &    32.28 &  2.1 $\pm$ 0.2 &  2.94 &      True &   True     \\
        SDSS J024929.18-002104.3 & 2549421 &   42.3716 &   -0.3512 &  1.4302 &   9.06 $\pm$ 0.11 &  45.61 $\pm$ 0.00 &     g &  SE &         \cite{Yang} &       \cite{Shen11} &       \cite{Shen11} &     7433 &    40.84 &  2.4 $\pm$ 0.5 &  1.23 &      True &   True     \\
        SDSS J025333.55+001634.3 & 2567252 &   43.3898 &    0.2762 &  1.4643 &   8.53 $\pm$ 0.35 &  44.81 $\pm$ 0.01 &     g &  SE &         \cite{Yang} &       \cite{Shen11} &       \cite{Shen11} &     7464 &    31.63 &  2.4 $\pm$ 0.3 &  2.25 &      True &   True     \\
        SDSS J024646.75-001220.5 & 2527604 &   41.6948 &   -0.2057 &  0.5636 &   8.31 $\pm$ 0.07 &  44.42 $\pm$ 0.05 &     g &  SE &         \cite{Yang} &       \cite{Shen11} &       \cite{Shen11} &     7458 &    33.90 &  2.4 $\pm$ 0.3 &  3.40 &      True &   True     \\
        SDSS J024028.10-005606.0 & 2591352 &   40.1171 &   -0.9350 &  1.0303 &   8.17 $\pm$ 0.41 &  44.70 $\pm$ 0.01 &     g &  SE &         \cite{Yang} &       \cite{Shen11} &       \cite{Shen11} &     7465 &    37.51 &  2.1 $\pm$ 0.2 &  2.40 &      True &   True     \\
        SDSS J024840.99-001228.8 & 2556507 &   42.1708 &   -0.2080 &  1.1992 &   8.91 $\pm$ 0.09 &  45.48 $\pm$ 0.00 &     g &  SE &         \cite{Yang} &       \cite{Shen11} &       \cite{Shen11} &     7458 &    23.60 &  2.5 $\pm$ 0.5 &  3.88 &      True &   True     \\
      KIC 006932990 & Zw 229-015 &  286.3582 &   42.4611 &  0.0273 &   7.00 $\pm$ 0.13 &  42.85 $\pm$ 0.00 &     V &  RM &      \cite{Barth11} &      \cite{Barth11} &      \cite{Barth11} &      436 &     1.70 &  1.5 $\pm$ 0.6 &  4.12 &      True &   True     \\
       NGC 4051 & NGC 4051 &  181.0434 &   44.5442 &  0.0023 &   5.67 $\pm$ 0.47 &  41.96 $\pm$ 0.19 &     V &  RM &  \cite{Fausnaugh17} &  \cite{Fausnaugh17} &  \cite{Fausnaugh17} &      219 &     0.82 &  1.1 $\pm$ 0.3 &  1.23 &      True &   True     \\
     PG J213227.82+100819.3 & PG 2130+099 &  323.1159 &   10.1387 &  0.0630 &   6.99 $\pm$ 0.10 &  44.40 $\pm$ 0.00 &     V &  RM &         \cite{Hu20} &         \cite{Hu20} &         \cite{Hu20} &      573 &     5.21 &  1.5 $\pm$ 0.4 &  9.61 &      True &   True     \\
     PG J135315.84+634545.7 & PG 1351+640 &  208.3160 &   63.7627 &  0.0870 &   7.66 $\pm$ 0.35 &  44.74 $\pm$ 0.04 &     B &  RM &      \cite{Kaspi00} &      \cite{Kaspi00} &      \cite{Kaspi00} &     2640 &    40.01 &  2.3 $\pm$ 0.3 &  3.12 &      True &   True     \\
        Mrk 142 & Mrk 142 &  156.3803 &   51.6764 &  0.0450 &   6.34 $\pm$ 0.23 &  43.31 $\pm$ 0.00 &     g &  RM &    \cite{Cackett20} &    \cite{Cackett20} &    \cite{Cackett20} &      231 &     0.64 &  1.2 $\pm$ 0.4 &  3.37 &      True &   True     \\
       NGC 3227 & NGC 3227 &  155.8912 &   19.8565 &  0.0038 &   6.66 $\pm$ 0.24 &  42.48 $\pm$ 0.00 &     V &  RM &    \cite{De_Rosa18} &    \cite{De_Rosa18} &    \cite{De_Rosa18} &      192 &     1.35 &  1.2 $\pm$ 0.4 &  1.96 &      True &   True     \\
        SDSS J153425.58+040806.7 & RGG 123 &  233.6066 &    4.1352 &  0.0395 &   5.10 $\pm$ 0.30 &  41.49 $\pm$ 0.00 &     r &  SE &          \cite{ZTF} &          \cite{RGG} &          \cite{RGG} &      528 &     3.77 &  1.5 $\pm$ 0.5 &  2.07 &      True &   True     \\
        SDSS J160531.85+174826.3 & RGG 127 &  241.3827 &   17.8073 &  0.0317 &   5.20 $\pm$ 0.30 &  41.17 $\pm$ 0.00 &     r &  SE &          \cite{ZTF} &          \cite{RGG} &          \cite{RGG} &      549 &     3.86 &  0.9 $\pm$ 0.2 &  1.19 &      True &   True     \\
       NGC 4253 & NGC 4253 &  184.6143 &   29.8113 &  0.0129 &   6.82 $\pm$ 0.05 &  42.51 $\pm$ 0.13 &     r &  RM &          \cite{ZTF} &            \cite{Bentz} &        \cite{Bentz} &      632 &     1.20 &  1.7 $\pm$ 0.7 &  1.96 &      True &   True \\
    PG J084742.46+344504.3 & PG 0844+349 &  131.9269 &   34.7512 &  0.0640 &   7.86 $\pm$ 0.19 &  44.24 $\pm$ 0.04 &     r &  RM &          \cite{ZTF} &      \cite{Kaspi00} &        \cite{Bentz} &      632 &     6.08 &  1.6 $\pm$ 0.5 &  1.98 &      True &   True     \\
        Mrk 50  & Mrk 50 &  185.8506 &    2.6790 &  0.0234 &   7.42 $\pm$ 0.06 &  42.73 $\pm$ 0.00 &     r &  RM &          \cite{ZTF} &         \cite{Le20} &        \cite{Bentz} &      639 &     7.02 &  1.5 $\pm$ 0.7 &  3.78 &      True &   True     \\
      Mrk 1044  & Mrk 1044 &   37.5230 &   -8.9981 &  0.0165 &   6.45 $\pm$ 0.12 &  42.67 $\pm$ 0.09 &     r &  RM &          \cite{ZTF} &            \cite{Bentz} &         \cite{Du14} &      517 &     3.80 &  1.1 $\pm$ 0.3 &  1.79 &      True &   True \\
        Arp 151 & Arp 151 &  171.4006 &   54.3825 &  0.0211 &   6.67 $\pm$ 0.05 &  42.48 $\pm$ 0.11 &     r &  RM &          \cite{ZTF} &            \cite{Bentz} &        \cite{Bentz} &      632 &     3.38 &  1.4 $\pm$ 0.3 &  1.58 &      True &   True \\
        Mrk 817 & Mrk 817 &  219.0920 &   58.7943 &  0.0315 &   7.59 $\pm$ 0.07 &  43.73 $\pm$ 0.05 &     r &  RM &          \cite{ZTF} &            \cite{Bentz} &        \cite{Bentz} &      640 &     2.04 &  1.5 $\pm$ 0.3 &  1.88 &      True &   True \\
        Mrk 335 & Mrk 335 &    1.5814 &   20.2029 &  0.0258 &   7.23 $\pm$ 0.04 &  43.68 $\pm$ 0.06 &     r &  RM &          \cite{ZTF} &            \cite{Bentz} &        \cite{Bentz} &      578 &     3.68 &  1.7 $\pm$ 0.6 &  1.55 &      True &   True \\
        \hline
\end{longtable}}

\bigskip
\end{landscape}
\restoregeometry

% do this here to save a page?
\clearpage
{\small
\noindent Caption for Data S1: \textbf{Summary data for the full AGN sample.} Same format as Table S1 but for all objects in the full AGN sample. A few objects with duplicate light curves were included in the full sample.
}


\begin{thebibliography}{61}

\bibitem{SSD}Shakura, N.~I., \& Sunyaev, R.~A.\ Black holes in binary systems. Observational appearance. {\em \aap}\ \textbf{24}, 337 (1973).

\bibitem{Shields}Shields, G.~A.\ Thermal continuum from accretion disks in quasars. {\em Nature} \textbf{272}, 706 (1978).

\bibitem{Sun_Malkan}Sun, W.H., \& Malkan, M.~A.\ Fitting Improved Accretion Disk Models to the Multiwavelength Continua of Quasars and Active Galactic Nuclei. {\em \apj} \textbf{346}, 68 (1989).

\bibitem{Morgan10}{Morgan}, C.~W., et~al.\ The Quasar Accretion Disk Size-Black Hole Mass Relation. {\em \apj} \textbf{712}, 1129 (2010).

\bibitem{Morgan18}{Morgan}, C.~W., et~al.\ Accretion Disk Size Measurement and Time Delays in the Lensed Quasar WFI 2033-4723. {\em \apj} \textbf{869}, 106 (2018).

\bibitem{Sergeev05}{Sergeev}, S.~G., et~al.\ Lag-Luminosity Relationship for Interband Lags between Variations in B, V, R, and I Bands in Active Galactic Nuclei. {\em \apj} \textbf{622}, 129 (2005).

\bibitem{Cackett07}Cackett, E.~M., Horne, K., \& Winkler, H. Testing thermal reprocessing in active galactic nuclei accretion discs. {\em \mnras} \textbf{380}, 669 (2007).

\bibitem{Fausnaugh}Fausnaugh, M.~M., et~al.\ Space Telescope and Optical Reverberation Mapping Project. III. Optical Continuum Emission and Broadband Time Delays in NGC 5548. {\em \apj} \textbf{821}, 56 (2016).

\bibitem{Edelson17}Edelson, R., et~al.\ Swift Monitoring of NGC 4151: Evidence for a Second X-Ray/UV Reprocessing. {\em \apj} \textbf{840}, 41 (2017).

\bibitem{Ulrich97}Ulrich, M.-H., Maraschi, L., \& Urry, C.~M.\ Variability of Active Galactic Nuclei. {\em \araa} \textbf{35} 445 (1997).

\bibitem{Padovani17}Padovani, P., et~al.\ Active galactic nuclei: what's in a name? {\em Astronomy and Astrophysics Review} \textbf{25}, 2 (2017).

\bibitem{kelly09}{Kelly}, B.~C., {Bechtold}, J., \& {Siemiginowska}, A.\ Are the Variations in Quasar Optical Flux Driven by Thermal Fluctuations? {\em \apj} \textbf{698}, 895 (2009).

\bibitem{koz10}{Koz{\l}owski}, S., et~al.\ Quantifying Quasar Variability as Part of a General Approach to Classifying Continuously Varying Sources. {\em \apj} \textbf{708}, 927 (2010).

\bibitem{MacLeod10}{MacLeod}, C.~L., et~al.\ Modeling the Time Variability of SDSS Stripe 82 Quasars as a Damped Random Walk. {\em \apj}\ \textbf{721}, 1014 (2010).

\bibitem{Suberlak21}{Suberlak}, K.~L., {Ivezi{\'c}}, {\v{Z}}., \& {MacLeod}, C.\ Improving Damped Random Walk Parameters for SDSS Stripe 82 Quasars with Pan-STARRS1. {\em \apj}\ \textbf{907}, 96 (2021)

\bibitem{Zu13}{Zu}, Y., {Kochanek}, C.~S., {Koz{\l}owski}, S., \& {Udalski}, A.\ Is Quasar Optical Variability a Damped Random Walk? {\em \apj} \textbf{765}, 106 (2013).

\bibitem{Simm16}{Simm}, T., et~al.\ Pan-STARRS1 variability of XMM-COSMOS AGN. II. Physical correlations and power spectrum analysis. {\em \aap} \textbf{585}, A129 (2016).

\bibitem{Mushotzky}{Mushotzky}, R.~F., {Edelson}, R., {Baumgartner}, W., \& {Gandhi}, P.\ Kepler Observations of Rapid Optical Variability in Active Galactic Nuclei. {\em \apjl} \textbf{743}, L12 (2011).

\bibitem{Collier_Peterson01}Collier, S., \&  Peterson, B.~M.\ Characteristic Ultraviolet/Optical Timescales in Active Galactic Nuclei. {\em \apj}\ \textbf{555}, 775 (2001).

\bibitem{koz17}{Koz{\l}owski}, S.\ Limitations on the recovery of the true AGN variability parameters using damped random walk modeling. {\em \aap} \textbf{597}, A128 (2017).

\bibitem{methods}Materials and methods are available as supplementary materials.

\bibitem{Peterson}{Peterson}, B.~M.\ Measuring the Masses of Supermassive Black Holes. {\em Space Science Reviews} \textbf{183}, 253 (2014).

\bibitem{Shen13}Shen, Y.\ The mass of quasars. {\em Bulletin of the Astronomical Society of India}\ \textbf{41}, 61 (2013).

\bibitem{Kollmeier06}Kollmeier, J.~A., et~al.\ Black Hole Masses and Eddington Ratios at $0.3< z < 4$. {\em \apj} \textbf{648}, 128 (2006).

\bibitem{Shen11} {Shen}, Y., {et~al.}\ A catalog of quasar properties from Sloan Digital Sky Survey data release 7. {\em \apjs}\ \textbf{194}, 45 (2011).

\bibitem{Scaringi15}Scaringi, S., et~al.\ Accretion-induced variability links young stellar objects, white dwarfs, and black holes. {\em Science Advances} \textbf{1}, e1500686 (2015).

\bibitem{Done}Done, C., Gierli{\'n}ski, M., \& {Kubota}, A.\ Modelling the behaviour of accretion flows in X-ray binaries. Everything you always wanted to know about accretion but were afraid to ask. {\em Astronomy and Astrophysics Review}\ \textbf{15}, 1-66 (2007).

\bibitem{MRI}Balbus, S.~A., \& Hawley, J.~F.\ Instability, turbulence, and enhanced transport in accretion disks. {\em Reviews of Modern Physics} \textbf{70}, 1 (1998).

\bibitem{xray}{Gonz{\'a}lez-Mart{\'\i}n}, O., \& {Vaughan}, S.\ X-ray variability of 104 active galactic nuclei. XMM-Newton power-spectrum density profiles. {\em \aap} \textbf{544}, A80 (2012).

\bibitem{McHardy06}McHardy, I.~M., et~al.\ Active galactic nuclei as scaled-up Galactic black holes. {\em Nature} \textbf{444}, 730 (2006).

\bibitem{Kording07}K{\"o}rding, E.~G., et~al.\ The variability plane of accreting compact objects. {\em \mnras}\ \textbf{380}, 301-310 (2007). 

\bibitem{Dexter}Dexter, J., {Agol}, E.\ Quasar Accretion Disks are Strongly Inhomogeneous. {\em \apj}\ \textbf{727}, L24 (2011).

\bibitem{Char}Sun, M., \etal\ Corona-heated Accretion-disk Reprocessing: A Physical Model to Decipher the Melody of AGN UV/Optical Twinkling. {\em \apj}\ \textbf{891}, 178 (2020).

\bibitem{Zenodo}Burke, C. J., \etal\ A characteristic optical variability timescale in astrophysical accretion disks [Data set], version 1, {Zenodo,}\ \url{http://doi.org/10.5281/zenodo.4914484}.

% additional references in the supplementary materials

\bibitem{Bentz}{Bentz}, M.~C., \& {Katz} S.\ The AGN Black Hole Mass Database. {\em Publications of the Astronomical Society of the Pacific}\ \textbf{127}, 67 (2015).

\bibitem{Geha03}{Geha}, M., et~al.\ Variability-Selected Quasars in MACHO Project Magellanic Cloud Fields. {\em \aj}\ \textbf{125}, 1 (2003).

\bibitem{Yang}{Yang}, Q., et~al.\ Dust Reverberation Mapping in Distant Quasars from Optical and Mid-Infrared Imaging Surveys. {\em \apj}\ \textbf{900}, 58 (2020).

\bibitem{Giveon99}{Giveon}, U., et~al.\ Long-term optical variability properties of the Palomar-Green quasars. {\em \mnras}\ \textbf{306}, 3 (1999).

\bibitem{Walsh09}{Walsh}, J.~L., et~al.\ The Lick AGN Monitoring Project: Photometric Light Curves and Optical Variability Characteristics. {\em \apjs}\ \textbf{185}, 156 (2009).

\bibitem{RGG}{Reines}, A.~E., {Greene}, J.~M., {Geha}, M.\ Dwarf Galaxies with Optical Signatures of Active Massive Black Holes. {\em \apj}\ \textbf{775}, 116 (2013).

\bibitem{Chilingarian18}{Chilingarian}, I.~V., et~al.\ A Population of Bona Fide Intermediate-mass Black Holes Identified as Low-luminosity Active Galactic Nuclei. {\em \apj}\ \textbf{863}, 1 (2018).

\bibitem{Guo}{Guo}, H., et~al.\ Dark Energy Survey Identification of A Low-Mass Active Galactic Nucleus at Redshift 0.823 from Optical Variability. {\em \mnras}\ \textbf{496}, 3636 (2020).

\bibitem{Cann20}{Cann}, J.~M., et~al.\ Multiwavelength Observations of SDSS J105621.45+313822.1, a Broad-line, Low-metallicity AGN. {\em \apj}\ \textbf{895}, 147 (2020).

\bibitem{Woo}{Woo}, J., et~al.\ A 10,000-solar-mass black hole in the nucleus of a bulgeless dwarf galaxy. {\em Nature Astronomy}\ \textbf{3}, 755 (2019).

\bibitem{TESS}{Ricker}, G.~R., et~al.\ Transiting Exoplanet Survey Satellite (TESS). {\em Journal of Astronomical Telescopes, Instruments, and Systems}\ \textbf{1(1)}, 014003 (2014).

\bibitem{Burke20}{Burke}, C.~J., et~al.\ Optical Variability of the Dwarf AGN NGC 4395 from the Transiting Exoplanet Survey Satellite. {\em \apj}\ \textbf{899}, 136 (2020).

\bibitem{Du18}{Du}, P., et~al.\ Supermassive Black Holes with High Accretion Rates in Active Galactic Nuclei. IX. 10 New Observations of Reverberation Mapping and Shortened H$\beta$ Lags. {\em \apj}\ \textbf{856}, 6 (2018).

\bibitem{Bentz14}{Bentz}, M.~C., et~al.\ The Mass of the Central Black Hole in the Nearby Seyfert Galaxy NGC 5273. {\em \apj}\ \textbf{796}, 8 (2014).

\bibitem{Bentz16a}{Bentz}, M.~C., et~al.\ A Reverberation-Based Black Hole Mass for MCG-06-30-15. {\em \apj}\ \textbf{830}, 136 (2016).

\bibitem{Bentz16b}{Bentz}, M.~C., et~al.\ The Host Galaxy of the Low Mass Black Hole in UGC 06728. {\em \apj}\ \textbf{831}, 2 (2016).

\bibitem{Pei14}{Pei}, L., et~al.\ Reverberation mapping of the \emph{Kepler} field AGN KA1858+4850. {\em \apj}\ \textbf{795}, 38 (2014).

\bibitem{Barth11}{Barth}, A.~J., et~al.\ Broad-Line Reverberation in the \emph{Kepler}-Field Seyfert Galaxy Zw 229-015. {\em \apj}\ \textbf{732}, 121 (2011).

\bibitem{Peterson14}{Peterson}, B.~M., et~al.\ Reverberation Mapping of the Seyfert 1 Galaxy NGC 7469. {\em \apj}\ \textbf{795}, 149 (2014).

\bibitem{Lu16}{Lu}, K., et~al.\ Reverberation Mapping of the Broad-line Region in NGC 5548: Evidence for Radiation Pressure?. {\em \apj}\ \textbf{827}, 118 (2016).

\bibitem{Hu20}{Hu}, C., et~al.\ Broad-line Region of the Quasar PG 2130+099 from a Two-year Reverberation Mapping Campaign with High Cadence. {\em \apj}\ \textbf{890}, 71 (2020).

\bibitem{Fausnaugh17}{Fausnaugh}, M.~M., et~al.\ Reverberation Mapping of Optical Emission Lines in Five Active Galaxies. {\em \apj}\ \textbf{840}, 97 (2017).

\bibitem{Bentz09}{Bentz}, M.~C., et~al.\ The Lick AGN Monitoring Project: Broad-line Region Radii and Black Hole Masses from Reverberation Mapping of H$\beta$. {\em \apj}\ \textbf{705}, 199 (2009).

\bibitem{Kaspi00}{Kaspi}, S., et~al.\ Reverberation Measurements for 17 Quasars and the Size-Mass-Luminosity Relations in Active Galactic Nuclei. {\em \apj}\ \textbf{533}, 631 (2000).

\bibitem{Cackett20}{Cackett}, E.~M., et~al.\ Supermassive Black Holes with High Accretion Rates in Active Galactic Nuclei. XI. Accretion Disk Reverberation Mapping of Mrk 142. {\em \apj}\ \textbf{896}, 1 (2020).

\bibitem{Denney10}{Denney}, K.~D., et~al.\ Reverberation Mapping Measurements of Black Hole Masses in Six Local Seyfert Galaxies. {\em \apj}\ \textbf{721}, 715 (2010).

\bibitem{De_Rosa18}{De Rosa}, G., et~al.\ Velocity-resolved Reverberation Mapping of Five Bright Seyfert 1 Galaxies. {\em \apj}\ \textbf{866}, 133 (2018).

\bibitem{Du14}{Du}, P., et~al.\ Supermassive Black Holes with High Accretion Rates in Active Galactic Nuclei. IV. H$\beta$ Time Lags and Implications for Super-Eddington Accretion. {\em \apj}\ \textbf{806}, 22 (2015).

\bibitem{Pancoast19}{Pancoast}, P., et~al.\ The Lick AGN Monitoring Project 2011: Photometric Light Curves. {\em \apj}\ \textbf{871}, 108 (2019).

\bibitem{peterson04}{Peterson}, B.~M., et~al.\ Central Masses and Broad-Line Region Sizes of Active Galactic Nuclei. II. A Homogeneous Analysis of a Large Reverberation-Mapping Database. {\em \apj}\ \textbf{613}, 682 (2004).

\bibitem{williams2018}{Williams}, P.~R., et~al.\ The Lick AGN Monitoring Project 2011: Dynamical Modeling of the Broad-line Region. {\em \apj}\ \textbf{866}, 75 (2018).

\bibitem{DES}{Dark Energy Survey Collaboration: Abbot}, A.~J., et~al.\ The Dark Energy Survey: more than dark energy – an overview. {\em \mnras}\ \textbf{460}, 1270 (2016).

\bibitem{ZTF}{Masci}, F.~J., et~al.\ The Zwicky Transient Facility: Data Processing, Products, and Archive. {\em Publications of the Astronomical Society of the Pacific}\ \textbf{131}, 995 (2018).

\bibitem{CRTS}{Drake}, A.~J., et~al.\ First Results from the Catalina Real-Time Transient Survey. {\em \apj}\ \textbf{696}, 870 (2009).

\bibitem{ASASSN}{Kochanek}, C.~S., et~al.\ The All-Sky Automated Survey for Supernovae (ASAS-SN) Light Curve Server v1.0. {\em Publications of the Astronomical Society of the Pacific}\ \textbf{129}, 104502 (2017).

\bibitem{Savic18}{Savi\'{c}}, D., et~al.\ AGN black hole mass estimates using polarization in broad emission lines. {\em \aap}\ \textbf{614}, A120 (2018).

\bibitem{Le20}{Le}, H.~A.~N., {Woo}, J., \& {Xue} Y.\ Calibrating Mg II-based black-hole mass estimators using Low-to-High-Luminosity Active Galactic Nuclei. {\em \apj}\ \textbf{901}, 35 (2020).

\bibitem{Du19}{Du}, P., \& {Wang}, J.\ The Radius–Luminosity Relationship Depends on Optical Spectra in Active Galactic Nuclei. {\em \apj}\ \textbf{886}, 42 (2019).

\bibitem{kelly14}{Kelly}, B.~C., et~al.\ Flexible and Scalable Methods for Quantifying Stochastic Variability in the Era of Massive Time-domain Astronomical Data Sets.\ {\em \apj}\ \textbf{788}, 33 (2014). 

\bibitem{Foreman-Mackey17}{Foreman-Mackey}, D., {Agol} E., {Ambikasaran} S., \& {Angus} R.\ Fast and Scalable Gaussian Process Modeling with Applications to Astronomical Time Series. {\em \aj}\ \textbf{154}, 220 (2017).

\bibitem{emcee}{Foreman-Mackey}, D., {Hogg} D.~W., {Lang} D., \& {Goodman} J.\ emcee: The MCMC Hammer. {\em Publications of the Astronomical Society of the Pacific}\ \textbf{125}, 306 (2013).

\bibitem{Lomb76}{Lomb}, N.~R.\ Least-squares frequency analysis of unequally spaced data. {\em Astrophysics and Space Science}\ \textbf{39}, 447 (1976).

\bibitem{Scargle82}{Scargle}, J.~D.\ Studies in astronomical time series analysis. II. Statistical aspects of spectral analysis of unevenly spaced data. {\em \apj}\ \textbf{263}, 835 (1982).

\bibitem{Uttley02}{Uttley}, P., {McHardy} I.~M., {Papadakis} I.~E.\ Measuring the broad-band power spectra of active galactic nuclei with RXTE. {\em \mnras}\ \textbf{332}, 231 (2002).

\bibitem{McHardy04}{McHardy}, I.~M. et~al.\ Combined long and short time-scale X-ray variability of NGC 4051 with \emph{RXTE} and \emph{XMM–Newton}. {\em \mnras}\ \textbf{348}, 783 (2004).

\bibitem{Kelly07}{Kelly}, B.~C.\ Some Aspects of Measurement Error in Linear Regression of Astronomical Data. {\em \apj}\ \textbf{665}, 1489 (2007).

\bibitem{Davis_Laor}{Davis}, S.~W., \& {Laor}, A.\ The Radiative Efficiency of Accretion Flows in Individual Active Galactic Nuclei. {\em \apj} \textbf{728}, 98 (2011). 

\bibitem{Peterson_4395}Peterson, B.~M., et~al.\ Multiwavelength Monitoring of the Dwarf Seyfert 1 Galaxy NGC 4395. I. A Reverberation-based Measurement of the Black Hole Mass. {\em \apj} \textbf{632}, 799 (2005).

\bibitem{denbrok}{den Brok}, M., et~al.\ Measuring the Mass of the Central Black Hole in the Bulgeless Galaxy NGC 4395 from Gas Dynamical Modeling. {\em \apj}\ \textbf{809}, 101 (2015).

\bibitem{Rucinski18}{Rucinski}, S.~M., et~al.\ Light-curve Instabilities of {\ensuremath{\beta}} Lyrae Observed by the BRITE Satellites. {\em \aj}\ \textbf{156}, 12 (2018)

\bibitem{JIA20}Jiang, Y.-F., \& Blaes, O. Opacity-driven Convection and
Variability in Accretion Disks around Supermassive Black Holes. {\em \apj} \textbf{900}, 25 (2020).

\bibitem{SCE18}Scepi, N., Lesure, G., Dubus, G., \& Flock, M.\ Impact of
Convection and Resistivity on Angular Momentum Transport in Dwarf Novae. {\em \aap} \textbf{609}, A77 (2018).

\bibitem{HIR14}Hirose, S., et~al.\ Convection Causes Enhanced Magnetic Turbulence in Accretion Disks in Outburst. {\em \apj} \textbf{787}, 1 (2014).

\bibitem{Hawley13}Hawley, J.~F., Richers, S.~A., Guan, X., \& Krolik, J.~H.\ Testing Convergence for Global Accretion Disks.  {\em \apj}, \textbf{772}, 102 (2013).

\bibitem{PRO00}Proga, D., Stone, J. M., \& Kallman, T. R.\ Dynamics of Line-Driven Disk Winds in Active Galactic Nuclei.  {\em \apj} \textbf{543}, 696 (2000).

\bibitem{NIX19}Nixon, C. J., \& Pringle, J. E.\ What is Wrong with Steady Accretion Discs? {\em \aap} \textbf{628}, A121 (2019). 

% Ref only in Figure caption
\bibitem{Timmer_Kong95}{Timmer}, J. \& {K\"{o}ng}, M.\ On generating power law noise. {\em \aap}\ \textbf{300}, 707 (1995).

% Refs only in Table 1
\bibitem{Cho}Cho, H., et~al.\ Variability and the Size–Luminosity Relation of the Intermediate-mass AGN in NGC 4395. {\em \apj} \textbf{892}, 93 (2020).

\end{thebibliography}
\end{document}